\documentclass[prd,aps,floatfix,superscriptaddress,nofootinbib,twocolumn,10pt,English]{revtex4-1}
\usepackage{amssymb,amsmath}
\usepackage{graphicx}
\usepackage{color}
\usepackage[colorlinks=true]{hyperref}

\def\be{\begin{equation}}
\def\ee{\end{equation}  }
\def\bea{\begin{eqnarray}}
\def\eea{\end{eqnarray}  }

\begin{document}          

\title{Black Hole Formation from the Collision of Plane-Fronted Gravitational Waves}
{
\author{Frans Pretorius}
\affiliation{Department of Physics, Princeton University, Princeton, New Jersey 08544, USA.}
\affiliation{CIFAR, Cosmology \& Gravity Program, Toronto, Ontario M5G 1Z8, Canada}
\author{William E. East}
\affiliation{Perimeter Institute for Theoretical Physics, Waterloo, Ontario N2L 2Y5, Canada}

\begin{abstract}
This paper introduces a new effort to study the collision of plane-fronted 
gravitational waves in four dimensional, asymptotically flat spacetime,
using numerical solutions of the Einstein equations. 
The pure vacuum problem
requires singular, Aichelburg-Sexl-type sources to achieve 
finite energy solutions, which are problematic to treat both mathematically and numerically. 
Instead then, we use null (massless) particles to source nontrivial geometry within the initial 
wave fronts.  The main purposes of this paper
are to (a) motivate the problem, (b) introduce methods for numerically solving the Einstein equations
coupled to distributions of collisionless massless or massive particles,
and (c) present a first result on the formation
of black holes in the head-on collision of axisymmetric distributions
of null particles. Regarding the last-named, initial conditions are chosen so that a black hole 
forms promptly, with essentially no matter escaping the collision. This can be interpreted
as approaching the ultrarelativistic collision problem from within an
infinite boost limit, but where the matter distribution is spread out, and thus
nonsingular. 
We find results that are consistent with earlier
perturbative calculations of the collision of Aichelburg-Sexl singularities, as well
as numerical studies of the high-speed collision of boson stars, black holes, and fluid
stars: a black hole is formed containing most of the energy of the
spacetime, with the remaining $15\pm1\%$ of the initial energy radiated
away as gravitational waves. 
The methods developed here could be relevant for other problems in strong-field gravity
and cosmology that involve particle distributions of matter. 
\end{abstract}

\maketitle

\section{Introduction}
As a theory of gravity, one aspect of general relativity that bares stark
contrast to Newton's theory is the nonlinear nature of the Einstein field 
equations. This imbues general relativity with a strong-field regime,
loosely defined as the class of solutions where the
nonlinearities of the theory are manifest, exhibiting qualitatively
different properties and phenomena than weak-field solutions. These include
black holes, cosmological solutions describing the global structure and evolution of the Universe, 
and gravitational collapse. The last-named refers to situations where, beginning
from nonsingular initial data (even weak-field data where linearity is satisfied
to good approximation early on), evolution eventually leads to the formation of some
kind of singularity in the structure of spacetime. 

Barring questions about the very early universe (the pre-inflationary epoch, or bounce in
cyclic models), and whether dark energy or dark matter might be due to a failure
of general relativity describing gravity on these scales, knowledge of strong-field
gravity relevant to the observable universe has become quite thorough over
the past several decades. This includes the geometries of neutron stars and 
black hole exteriors, the spacetime dynamics in mergers of such compact objects,
and knowledge of how the exterior geometry
settles to a stationary Kerr solution when gravitational collapse occurs (see, 
e.g.,~\cite{Shibata:2011jka,Fryer:2011zz,Faber:2012rw,Blanchet:2013haa,Lehner:2014asa,Paschalidis:2016vmz} 
for some review articles). Of course, many details of
these processes remain to be understood, and fundamental open questions
remain where matter plays an important or dominant role---core-collapse 
supernova, self-consistent models connecting binary neutron
star mergers to the host of observed/hypothesized electromagnetic counter parts,
how accreting black holes power observed jets, etc.

However, many other aspects of strong-field general relativity (of arguably less
astrophysical importance, but still of significant physical and mathematical interest)
remain poorly understood, despite the relevant open questions having
been identified decades ago. These include the interior structure of black holes
and the generic nature of singularities formed in gravitational collapse, 
critical phenomena at the threshold of black hole formation (in particular
in situations lacking symmetries), the classification and properties
of solutions with event horizons in higher dimensional spacetimes,
gravity in asymptotically anti-de Sitter (AdS) spacetimes, and the ultrarelativistic scattering problem
(see e.g.~\cite{Gundlach:2002sx,Berger:2002st,Emparan:2008eg,Dafermos:2008en,Chesler:2013lia,Choptuik:2015mma,Garfinkle:2016lcu,Dafermos:2017dbw,Barack:2018yly,Coley:2018mzr}).
The goal of this paper is to introduce a new research effort to tackle aspects
of this last-named problem.

As part of this effort, we develop methods for evolving distributions of
collisionless particles coupled to Einstein gravity. Though the main focus of
this work is the null-particle case, these same methods can be used for massive
particles, as we demonstrate in passing in this work. This could be relevant for
tackling a number of other problems in strong-field gravity, such as studies
of cosmic censorship~\cite{Shapiro:1991zza} or inhomogeneous cosmologies.
Recently, there has been interest in performing fully general-relativistic
calculations of structure formation
scenarios~\cite{Bentivegna:2015flc,Giblin:2015vwq,Giblin:2016mjp,Macpherson:2016ict,East:2017qmk,Macpherson:2018btl} to study effects not captured by standard Newtonian N-body calculations.
However, current approaches rely on fluid descriptions that break down in the
presence of multistream regions that arise, e.g., during halo formation, 
or use weak-field approximations to the Einstein equations~\cite{Adamek:2015eda}.

Before outlining the main content of the remainder of the paper, we describe the
ultrarelativistic scattering problem in more detail, giving some background, and a list of open questions we
would eventually like to address.

\subsection{The Ultrarelativistic scattering problem}
In the context of strong-field gravity, the ultrarelativistic scattering problem
refers to understanding the dynamics of spacetime following the
interaction of two distributions of energy, initially approaching each other
from opposite directions near or at the speed of light as measured in the center-of-momentum
frame. The source of energy could be pure spacetime itself, i.e. plane-fronted
gravitational waves or black holes (including Aichelburg-Sexl (AS) singularities in the
infinite boost limit), or some classical model of a particle, such as boson stars, fluid stars,
or even black holes. Two seminal works in this area were initiated
by Khan and Penrose~\cite{Khan71}, and Penrose~\cite{Penrose74} almost 50 years ago,
essentially addressing two opposite extremes in the landscape
of the scattering problem : collision of waves of infinite transverse extent, and
the collision of point particles. 

\subsubsection{Collision of planar gravitational waves}
Khan and Penrose~\cite{Khan71} studied the collision of two infinite, plane-fronted gravitational
waves, and found the interaction always resulted in a naked curvature singularity, regardless
of how weak (in terms of curvature) each individual wave was. The formation of the singularity
can be understood as the result of the focusing of the geometry of one wave front by the
other, and vice versa. The weaker the initial curvature is, the longer it takes
to focus to a singularity, though it always does. One can argue that formation
of a singularity in this scenario is an artifact of both the perfect focusing
caused by the planar geometry of the wave front, as well as its infinite transverse extent
(hence the spacetime has infinite energy and is not asymptotically flat). Moreover,
even in the absence of any dynamics resulting from the collision of two such
wave fronts, a single wave, that is locally exactly Minkowski
spacetime on either side of the front, is not globally hyperbolic: any spacelike
hypersurface of the single wave geometry evolved forward in time using the Einstein equations will encounter
a Cauchy horizon~\cite{1965RvMP...37..215P}. In a sense, uniform plane-fronted
gravitational waves are infinitely powerful lenses, capable of focusing all of
Minkowski spacetime down to a point. We illustrate this more explicitly in
Sec.~\ref{plane_waves}. 

The open questions pertaining to this limit
of the ultrarelativistic scattering problem then relate to how these
``pathological'' outcomes might change if the apparent sources
of the pathology 
---a uniform  distribution and infinite total energy---
are removed. If the energy density in each wave has finite transverse
extent, one would expect the focusing to end within a transverse light-crossing time. If this
is longer than the time for the singularity to form, will a black hole eventually form to censor
the singularity? How do inhomogeneities in the energy density affect the evolution
of the wave-front post collision?\footnote{These questions could be of relevance
to certain bubble formation/collision scenarios in the early universe~\cite{Hawking:1982ga}. 
A bubble of true vacuum
nucleated in a false vacuum will expand, with the bubble wall gaining all the 
energy of the false vacuum swept up. The wall quickly becomes relativistic, and eventually
strongly self-gravitating. The wall will therefore begin acting as a strong lens focusing
matter it encounters, and conversely inhomogeneity in the matter will backreact to create inhomogeneity in
the bubble wall. If two such bubbles collide, in a region about the initial point of contact 
much smaller than the radius of curvature of each bubble, to good approximation
the collision could be treated as the collision of two plane-fronted waves.
The curvature in the walls around the collision is not expected
to be strong enough to focus to singularities before the nonplanar
nature of the walls influence the dynamics (if that were the case, the bubbles
would have individually collapsed to black holes prior to the collision(see e.g.~\cite{Deng:2018cxb})).
Though self-gravitating bubble collisions have been studied extensively 
before (see e.g.~\cite{Wainwright:2013lea,Wainwright:2014pta,Johnson:2015gma} and the references therein),
these studies have not included the effect of inhomogeneities, which would be interesting
to explore. See~\cite{Braden:2014cra,Braden:2015vza,Bond:2015zfa} for related work suggesting that domain walls (neglecting self-gravity) are unstable to perturbations.}

\subsubsection{Collision of point particles}
At the other end of the spectrum, Penrose~\cite{Penrose74} initiated the study
of the collision of plane-fronted waves sourced by singular, pointlike
distributions of energy. Such a plane wave geometry can be obtained by taking
the Schwarzschild solution with rest mass $m$, boosting it with Lorentz factor
$\gamma$, and considering the limit that $\gamma\rightarrow\infty$ and
$m\rightarrow 0$ while the energy $E=m \gamma$ remains fixed. This yields the
Aichelburg-Sexl solution, which can be considered to
describe the geometry of a null particle~\cite{Aichelburg:1970dh}.
The AS
geometry is Minkowski on either side of the shock front, with 
all the curvature confined to the transverse plane. The magnitude
of the gravitational wave as measured by a Newman-Penrose scalar
drops like $1/\rho^4$ with transverse distance $\rho$ from the 
origin,
and though not asymptotically flat in the strictest sense of the definition~\cite{1992PhRvD..46..658D}, 
the spacetime still approaches Minkowski at large $\rho$,
and does not suffer the infinite energy/focusing pathologies of the previously discussed
uniform plane wave solutions. This, however, comes at the expense of the origin now being
a naked, curvature singularity. Nevertheless, Penrose was able to show that an apparent horizon
exists in the spacetime formed by the superposition of two colliding AS shocks;
the exact solution to the causal future of the collision is unknown, but
this implies a black hole forms. Though an extreme example of relaxing the
infinite extent problems of uniform plane wave collisions, this is suggestive that the pathologies
with the latter scenario are indeed due to infinite extent, and not properties of the nonlinear
interaction in gravitational wave scattering. 

The study of the point particle limit was reinvigorated a couple of decades
ago following the realization that if extra dimensions exist, the true
Planck scale could be orders of magnitude lower than the scale naively
inferred from 4D dimensional analysis~\cite{1998PhLB..429..263A,1998PhLB..436..257A,1999PhRvL..83.3370R}.
It was argued that one ``natural'' magnitude for the Planck energy is $O({\rm TeV})$. This
renewed interest in the gravitational scattering problem because particle
collisions above the Planck energy are generically expected to form
black holes~\cite{1987PhLB..198...61T,1987PhLB..197...81A,1999hep.th....6038B}, 
so if the low-TeV Planck-scale scenario is true, the Large Hadron Collider,
as well as high energy cosmic ray collisions with the Earth's atmosphere, could
then form black holes~\cite{2001PhRvL..87p1602D,Giddings:2001bu,2002PhRvL..88b1303F}. No evidence
for this has been found to date; see e.g.~\cite{2018arXiv180506013C}.

At sufficiently high energies in particle collisions, the black holes that form would
be large enough to censor any details of the collisions, and hence it is conjectured
that in the ultrarelativistic limit ``matter does not matter''; i.e., gravity
dominates the interaction, and moreover, the geometry of each boosted particle
is dominated by its kinetic energy, hence the ultrarelativistic limit
is uniquely captured by the collision of two AS shock waves.
This argument certainly makes intuitive sense, though from a geometric
perspective, given the rather stark differences between the geometries of large-but-finite boost
timelike compact objects and the singular, plane-fronted
AS shockwave solution\footnote{For example, all polynomial invariant scalars
of the Riemann tensor, such as the Kretschmann scalar, vanish for null, plane-fronted gravitational
wave spacetimes~\cite{Kramer80}, including the AS spacetime.}, it
would be rather remarkable if this conjecture were true.
Nevertheless, evidence has been
gathered in its favor from simulations of ultrarelativistic boson
star~\cite{Choptuik:2009ww}, fluid star~\cite{East:2012mb}, and black hole
collisions~\cite{Sperhake:2008ga}: for collisions between material objects,
black holes do form above a threshold roughly in line with
hoop-conjecture arguments~\cite{thorne_hoop}, and in all cases the
gravitational wave energy emission in head-on collisions, extrapolated
to infinite boosts, agrees with perturbative calculations
of that produced in the collision of two AS waves~\cite{1992PhRvD..46..694D}
(the last-named finding $\sim16\%$ of the initial mass of the spacetime being radiated).

Many open questions remain here. Regarding the apparent limiting behavior
of large boosts, one challenge to establish the connection with AS is that detailed
comparisons will require explicit solutions, and it is unclear
how to deal with the AS naked singularity, in particular in a numerical solution 
scheme.
The approach we propose here is to begin {\em at} the infinite boost limit, i.e. with null plane-fronted
waves, and replace the singularity with a smooth null matter source (such solutions
are sometimes called gyratons, the first examples of which were discovered
by Bonnor~\cite{1969CMaPh..13..163B} and Peres~\cite{1959PhRvL...3..571P,1960PhRv..118.1105P}).
This would also allow one to address how ``stable'' the AS singularity is in the first place,
by studying the stability of any family of regular spacetimes that approach the AS spacetime in some continuous limit.
One conceivable outcome of instability is that perturbations generically lead to black hole formation.

Not much is known about the dynamics of off-axis collisions. Perturbative calculations
of large impact parameter scattering suggests the ultrarelativistic two body
problem is significantly simpler than the more astrophysical, rest-mass dominated
regime~\cite{Galley:2013eba,2016CQGra..33l5012G}. This is consistent with studies that indicate
some of the leading order physics in these interactions, even black hole formation,
can be captured by appealing to geodesic motion on a single, relevant background 
geometry~\cite{Kaloper:2007pb,East:2012mb}. Another result from perturbative
calculations of large impact parameter deflections is the radiation produced is highly beamed~\cite{Galley:2013eba}.
This, together with geodesic focusing of the 
emitted radiation, could explain the rather striking growth in black hole
mass noted during a moderate-boost ($\gamma\sim1.5$) grazing encounter simulation 
of two black holes~\cite{Sperhake:2012me}.
It would be interesting to explore these grazing encounters at much higher boosts.

A further set of questions relates to the threshold of black hole formation. In particular,
if critical phenomena~\cite{Choptuik:1992jv} is present, and if so, when tuning
to threshold, which critical solution is revealed: that of the underlying matter source,
or of vacuum gravity.

Of course, not all questions pertaining to the ultrarelativistic scattering problem
need to try to make a connection with one of the two extreme limits : point particle
scattering or infinite uniform plane-wave scattering. There is potentially a vast
landscape of interesting phenomenology in between, worthy of study in its own right.

\subsection{Outline of the remainder of the paper}

In Sec.~\ref{sec_form} we describe the formalism in more detail: the Einstein
equations coupled to null matter, certain properties of plane-fronted wave solutions,
and the similarities/differences between the pure vacuum versus matter sourced
cases. In Sec.~\ref{sec_code} we describe a new code designed to solve
the Einstein-collisionless particle system, in particular highlighting the issues required
to self-consistently and efficiently compute the 
stress-energy tensor summed over the contributions from the discrete 
collection of particles. In Sec.~\ref{sec_results} we present results
from simulations of the head-on collision of two axisymmetric null-particle 
distributions, including convergence tests. The main result is that we find a spacetime
consistent with prior approaches to studying the AS collision limit: a black hole forms
containing $85.1\pm0.8\%$ of the initial energy of the spacetime,
with the rest escaping as gravitational radiation 
(modulo $<0.1\%$ in energy of the tail end of the null particle distribution
that did not fall into the black hole). Regarding the structure of the 
waves, we find it is highly beamed about the collision axis. Also,
the plane-fronted shock-like features of the initial geometries are 
trapped by the black hole, continually propagate about it,
albeit with an amplitude that exponentially decreases with time.
We conclude in Sec.~\ref{sec_conc} 
by mentioning some improvements to the code that would be needed before
all the topics discussed above can be addressed.
In the appendix we further validate the particle code, showing how it can
be applied to the massive particle case, by applying it to a simple inhomogeneous
cosmology setup. 

We use geometric units where Newton's constant $G$ and the speed of light $c$ are
set to unity.

\section{Formalism}\label{sec_form}

In this section we discussion the Einstein equations coupled
to a null fluid (Sec.~\ref{sec_null_fluid}); the problems
with homogeneous, plane-fronted gravitational wave spacetimes
(Sec.~\ref{plane_waves}); how these problems
can be resolved when using a null fluid as the source of the plane-wave
geometry (Sec.~\ref{reg_plane_waves}); and the issues of using a null
fluid to study colliding, regular plane-wave geometries, and how
null particle distributions can alleviate them (Sec.~\ref{sec_dust}).

\subsection{The Einstein field equations with a null fluid source}\label{sec_null_fluid}
We first consider the Einstein equations with a pressureless fluid as
a source: 
\bea\label{efe}
R_{ab}-\frac{1}{2}R g_{ab} &=& 8\pi T_{ab} \\
                           &\equiv& 8\pi \rho_e \ell_a \ell_b \nonumber,
\eea
where $R_{ab}$ is the Ricci tensor, $R$ the Ricci scalar, $g_{ab}$ the metric tensor,
$T_{ab}$ the stress-energy tensor, and $\rho_e$ is the energy density
flowing along the direction $\ell^a$. 
For a timelike pressureless fluid 
$\ell^a$ is the four velocity of the fluid (with $\ell^a \ell_a=-1$) and
$\rho_e$ has an unambiguous interpretation as the
rest-frame density of the fluid.
However, for a null fluid ($\ell^a \ell_a=0$), as we consider below, 
$\rho_e$ is the energy density
in {\em some} chosen frame. This frame then determines the normalization
of $\ell^a$, which is equivalent to fixing the affine parameter $\lambda$ in
the parametric representation of the null vector, $\ell^a=dx^a(\lambda)/d\lambda$
(or vice versa: a frame where the normalization of $\ell^a$ has been chosen can be thought
of as the frame defining the interpretation of $\rho_e$).

Later we will generalize this to a distribution of noninteracting null
particles, though for now the null fluid is more convenient to illustrate
plane-fronted wave solutions sourced by matter. Moreover, prior to the collision
of two such fronts, there is a simple one-to-one correspondence between
the fluid and particle solutions, and the former is more convenient to use to provide
initial data for the collision.

\subsection{Plane-fronted waves}\label{plane_waves}
Consider a plane-fronted wave (sometimes also referred to as a plane-parallel, or pp wave) 
traveling in the $+x$ direction---see Fig.~\ref{fig_1_pulse}.
The metric, in Brinkmann-like form, is
\be\label{brinkmann}
ds^2 = -du dv + B^2(u) \left[dy^2+dz^2\right] + H(u,y,z) du^2,
\ee
where $v=t+x$ is a null coordinate,
$u=t-x$ is a null coordinate when $H=0$, and $y,z$ are
the Cartesian-like coordinates transverse to the wave. There are several
different coordinate systems often used to represent 
plane-fronted metrics, all relatable to each other through coordinate
transformations (see e.g.~\cite{Kramer80}). Perhaps the most common one
sets $B(u)=1$, though we find it convenient to include this term here
for our discussion related to null-sourced waves, and to motivate the eventual
form of the metric we use to set initial conditions for numerical evolution
(in Sec.~\ref{id_specific} below). The only nontrivial Einstein equation (\ref{efe})
for this metric ansatz is
\be\label{efe11}
\frac{\nabla _2 H}{B^2} + \frac{4\ddot{B}}{B} + 8\pi\rho_e = 0,
\ee
where $\nabla_2\equiv\partial_{yy}+\partial_{zz}$ is the two dimensional flatspace
Laplacian, and the overdot ($\dot{\ }$) denotes $\partial_{u}$. With respect to orthonormal
basis vectors aligned with the coordinate directions, the only nonzero
Newman-Penrose scalar for this metric is $^{(x)}\Psi_4$, defined to measure
gravitational waves propagating in the $+x$ direction\footnote{Hence the $^{(x)}$-superscript,
to differentiate it from $^{(r)}\Psi_4$ shown in the results section, which is
calculated with a tetrad to measure radiation propagating in the radial
direction.}:
\be\label{psi4_brinkmann}
^{(x)}\Psi_4 = \frac{H_{,zz}-H_{,yy}}{2 B^2}-i \frac{H_{,yz}}{B^2},
\ee
where here and below we use a comma to denote the
ordinary (partial) derivative operator.

\begin{figure}
\vspace{0.1in}
\includegraphics[width=3.50in,draft=false,clip]{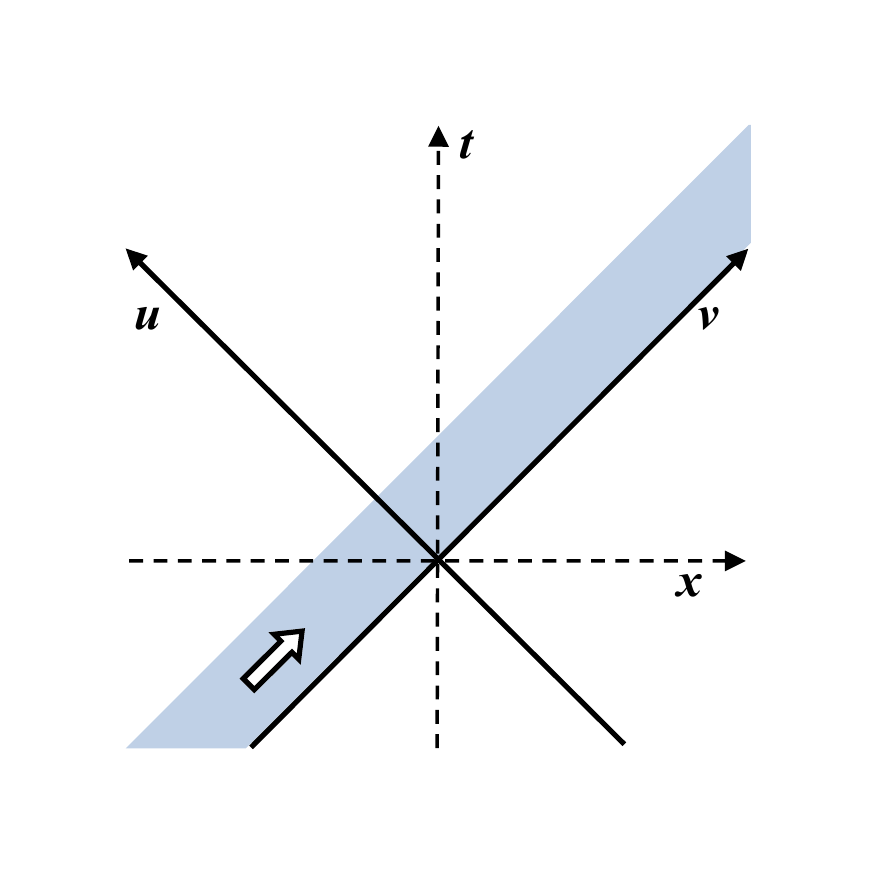}
\caption{
A spacetime diagram depicting a null-fronted plane wave (shaded
region) traveling
in the $+x$ direction. $u=t-x$ and $v=t+x$ are null coordinates; 
the two coordinates $y,z$ transverse to the wave are not shown. On either side
of the pulse $u<0$ and $u>\Delta u$, the geometry is that
of Minkowski spacetime.
}
\label{fig_1_pulse}
\end{figure}

To study pure gravitational wave spacetimes ($\rho_e=0$) it is most convenient
to set $B=1$ and $H(u,y,z)\equiv f(u) h(y,z)$; then the field equations
allow an arbitrary amplitude profile $f(u)$, with the transverse
structure constrained by $\nabla_2 h=0$, and it is manifest
that the spacetime is Minkowski away from the wave front (when $f=0$).
This form of the metric is adequate to capture the entire
class of vacuum, plane-fronted gravitational wave solutions~\cite{Kramer80},
and also highlights why these spacetimes are problematic to address some of 
the questions identified in the introduction: uniqueness properties
of $\nabla_2 h=0$ dictate the only way to obtain finite transverse
extent, or inhomogeneity in the transverse plane, is via
AS-like singularities, or boundary conditions at infinity.
The latter is essentially how inhomogeneities are introduced in
asymptotically AdS spacetimes, where gravitational
wave scattering is used to model heavy ion collisions using
gauge-gravity dualities (see e.g.~\cite{Chesler:2015fpa}), though this option is not available
in asymptotically flat spacetime. Before describing how null
dust can circumvent these problems, it is useful to consider
the homogeneous dust-sourced spacetimes, with $\rho_e=\rho_e(u)$. 

The simplest homogeneous dust solutions are those without any ``pure''
gravitational waves (i.e. all nontrivial curvature is in the trace-full
part of the Riemann tensor constrained by the Einstein
equations, with the Weyl tensor identically zero) : $H=0$, with 
$B(u)$ satisfying
\be\label{B_rho_u}
\ddot{B}+2\pi\rho_e B=0.
\ee
Suppose we have a matter pulse with support in $u\in[0,\Delta u]$ (Fig.~\ref{fig_1_pulse}),
and we choose initial conditions to (\ref{B_rho_u}) such that
$B(u<0)=1$; i.e., the metric [(\ref{brinkmann}) with $H=0$] is in standard Minkowski form
prior to passage of the wave. If $\rho_e$ is positive, $B(u>0)$ will begin to decrease as $u$ increases,
and always reach zero, either within the pulse for sufficiently high amplitude and/or wide pulses,
or sometime after the passage of the pulse. For example, for a constant density pulse
with $\rho_e=\rho_{e0}$ within $0\leq u \leq \Delta u$, $\rho_e=0$ otherwise,
the solution to (\ref{B_rho_u}) is
\bea\label{B_h_soln}
B(u)&=& 1, \ \ \ u<0 \nonumber\\
    &=&\cos(\omega_0 u), \ \ \ 0\leq u \leq \Delta u \nonumber\\
    &=&\cos(\omega_0 \Delta u) \nonumber\\
    & & - (u-\Delta u) (\omega_0\sin(\omega_0 \Delta u)), \ \ \ u>\Delta u,
\eea
with $\omega_0=\sqrt{2\pi\rho_{e0}}$. For the coordinate singularity $B=0$
to occur behind the pulse, we must have $\rho_{e0} < \pi/(8\Delta u^2$). To see that this is 
a coordinate singularity, consider the following coordinate transformation
($u$ is unchanged):
\bea\label{ctrans1}
y&=&\bar{y}/B(u),\nonumber\\
z&=&\bar{z}/B(u),\nonumber\\
v&=&\bar{v}+L(u)\left[\bar{y}^2+\bar{z}^2\right]
\eea
If we choose $L(u)=-\dot{B}/B$, then away from the matter where $\ddot{B}=0$, the above
transforms (\ref{brinkmann}) with $H=0$ to $ds^2=-du d\bar{v} + d\bar{y}^2 + d\bar{z}^2$. 

On the other hand, $B=0$ is more than a coordinate singularity. Consider the 
focusing caused by the passage of the wave on a 
set of timelike geodesics, initially at rest. In particular,
looking at the transverse coordinates of any such geodesic, if $y(u<0)=y_0$, $z(u<0)=z_0$, 
then the geodesic equation says $y(u)=y_0$, $z(u)=z_0$; i.e. in these
coordinates such geodesics remain at fixed transverse coordinate locations. 
However, from (\ref{brinkmann}) the proper transverse (geodesic) distance
between any pair of geodesics with coordinate separation ($\Delta y, \Delta z$) 
at some $u=v=\rm{constant}$ is $\ell_p=|B|\sqrt{\Delta y^2 + \Delta z^2}$. In other words,
the proper distance between {\em all} these timelike geodesics goes to zero
at $B=0$, when $u = u_s \equiv \Delta u + \cot(\omega_0 \Delta u)/\omega_0$.
In that sense the plane wave front 
is an infinitely powerful lens. 
Moreover, given that $(y,z)={\rm constant}$ curves are geodesics, and 
$(\partial/\partial y)^a$ and $(\partial/\partial z)^a$ are Killing vectors
of the spacetime, we can toroidally compactify the space by identifying
$y$ with $y+y_L$, and $z$ with $z+z_L$ for some constants $y_L,z_L$. The geodesic
focusing then tells us that the entire compactified transverse space is focused
to a point at $B=0$. This illustrates that it is more appropriate to think
of the plane wave front as focusing all of {\em spacetime},
and not merely a class of geodesics. Similarly, considering the regular
$(u,\bar{v},\bar{y},\bar{z})$ coordinate chart, except for 
where ($\bar{y}=0,\bar{z}=0$), the {\em entire} range $\bar{v}\in[-\infty,\infty]$
is mapped to $v=\infty$ when $u=u_s$. This demonstrates that these regular
coordinates cannot be used to specify complete, Cauchy data for the region
$u>u_s$.

\subsection{Regular, plane-fronted waves}\label{reg_plane_waves}
As mentioned above, to obtain vacuum plane-fronted gravitational waves
that have finite total energy and are asymptotically flat transverse to the
wave front requires singular sources. Using a null fluid
instead can remedy the problem. The form of the metric ansatz (\ref{brinkmann}) used
above is not ideal for numerical evolution, in
that for a localized source asymptotically $H\propto \ln(\rho)$, with $\rho\equiv\sqrt{y^2+z^2}$.
Moreover, the focusing can still lead to $B=0$ coordinate singularities near the
wake of the fluid. A better suited coordinate system can be found
appealing to the coordinate transformation (\ref{ctrans1}), which essentially
``unfocuses'' the metric following the wave (however unlike (\ref{ctrans1})
for a homogeneous wave, the finite wave has no Cauchy horizon problems).
For the axisymmetric collisions explored later, it will also be more convenient
to use cylindrical coordinates. We therefore use the following ansatz
for the metric and fluid source propagating along $u={\rm constant}$ characteristics, 
with the specific form of the functions chosen to simplify the resulting field equations:
\bea
ds^2&=&-du dv -8\pi f(u) h(\rho,\theta) du^2 \nonumber\\
    &\ & + 2\beta(u)q(\rho) du d\rho + d\rho^2 + \rho^2 d\theta^2,\label{metric_gen}\\
T^{ab} &=& \rho_e(u,\rho,\theta) \ell^a \ell^b,\label{cont_set}
\eea
with
\bea
\beta(u)&=& 4\pi \int_{-\infty}^u f(\tilde{u}) d\tilde{u},\label{beta_eqn}\\
\rho_e(u,\rho,\theta)&=&f(u)g(\rho,\theta),\label{rhoe_eqn}
\eea
and null vector normalized to $\ell^a=\sqrt{2}(\partial/\partial v)^a$.
With this ansatz, the one nontrivial Einstein equation is: 
\be\label{efe_cyl}
\nabla_2 h \equiv h_{,\rho\rho} + \frac{h_{,\rho}}{\rho} + \frac{h_{,\theta\theta}}{\rho^2} = g - q_{,\rho} - \frac{q}{\rho}
\ee
If we further decompose the transverse dependence of the energy density into cylindrical harmonics
\be\label{g_decomp}
g(\rho,\theta)\equiv g_0({\rho}) + \sum_{m=1}^{\infty} g_m(\rho) \cos(m\theta),
\ee
then we can use $q(\rho)$ to solve for the monopole contribution to (\ref{efe_cyl}):
\be\label{q_eqn}
q' + \frac{q}{\rho} = g_0,
\ee
where $'$ denotes differentiation with respect to $\rho$.
$h(\rho,\theta)$ then captures the metric response to a nonaxisymmetric
source, which can readily be solved via a similar decomposition
\be\label{h_eqn1}
h(\rho,\theta)\equiv \sum_{m=1}^{\infty} h_m(\rho) \cos(m\theta),
\ee}where for each $m$ the remaining portion of the field equation (\ref{efe_cyl})
reduces to the following ordinary differential equation (ODE)
\be\label{h_eqn2}
h_m^{''} + \frac{h_m^{'}}{\rho} - m^2 \frac{h_m}{\rho^2} = g_m.
\ee

When solving the above equations, 
we are free to choose the energy density via the
functions $f(u), g_0(\rho)$ and $g_m(\rho)$, and then solve for the remaining
metric functions $\beta(u),q(u),h(\rho,\theta)$ using (\ref{beta_eqn}),(\ref{q_eqn}),(\ref{h_eqn1}) and (\ref{h_eqn2}).
We will restrict the class of free initial data to that which is
regular in the limit $\rho\rightarrow 0$,
as well as an asymptotically flat spacetime in the limit $\rho\rightarrow \infty$.
In the limit $\rho\rightarrow 0$, regular solutions require
\bea
g_0(\rho) &=& \alpha_0 + O(\rho^2)\\
g_m(\rho) &=& \gamma_0 \rho^m + O(\rho^{m+2}),
\eea
where $\alpha_0$ and $\gamma_0$ are constants.
In the limit $\rho\rightarrow \infty$ we require $\rho_e\rightarrow 0$ sufficiently
rapidly to give the spacetime finite total mass. For the energy density
profiles, we either use compactly supported pulses ($\rho_e=0$ for $\rho>\rho_{\rm max}$),
or a Gaussian ($\rho_e\propto e^{-(\rho/\delta \rho)^2}$), the latter
which we use in the numerical evolution. Then, the metric (\ref{metric_gen})
asymptotes to Minkowski as $q\propto 1/\rho$, $h\propto 1/\rho^m$ (though we
have $h=0$ for the axisymmetric numerical solutions discussed later). 

We will use the Arnowitt-Deser-Misner (ADM) mass~\cite{Arnowitt:1962hi}
as a measure of the spacetime energy, integrated on a cylinder at (arbitrarily)
$t=0$, with $u=t-x,v=t+x$, centered about the pulse, and taking the size of the cylinder to $\infty$:
\bea\label{adm1}
M_{\rm ADM} &=& \frac{\beta(\infty)}{8}\bigg[ \int_0^\infty \left( 
           \frac{d q(\tilde{\rho})}{d\tilde{\rho}}\tilde{\rho} + q(\tilde{\rho})\right) \  d\tilde{\rho} \nonumber \\
        &\ & + \lim_{\rho\rightarrow \infty} \rho q(\rho)\bigg].
\eea
In reaching the above form for $M_{\rm ADM}$, we have assumed the pulse has finite extent in $u$, and
that $h\rightarrow 0$ at least as fast as $1/\rho$; thus the asymmetric piece
of the metric does not contribute to the ADM mass. We can simplify
the expression using (\ref{q_eqn}), giving
\be\label{adm2}
M_{\rm ADM} = \frac{\beta(\infty)}{8}\left[ \int_0^\infty \tilde{\rho} g_0(\tilde{\rho}) \  d\tilde{\rho} + \lim_{\rho\rightarrow \infty} \rho q(\rho)\right].
\ee
We are not aware of studies that have explored the validity of the ADM
mass measure for this class of spacetime in these coordinates, with the
exception of the AS solution itself~\cite{1998CQGra..15.3841A,2000CQGra..17.3645A}. 
In Sec.~\ref{id_specific} we show that
with the family of initial data we use, taking the AS limit with fixed $M_{\rm ADM}$ as defined
above does give the AS solution with mass parameter $M$ equal to $M_{\rm ADM}$. 
Interestingly though,
if we directly evaluate (\ref{adm2}) with the exact AS solution, we obtain $M/2$.
For the regular, limiting sequence of solutions there is equal contribution of $M/2$ from the integral
along the end cap of the cylinder (first term in (\ref{adm2})) and the integral
over the barrel of the cylinder (second term in (\ref{adm2})). However, the former
piece identically vanishes in vacuum. The AS solution
is singular on the axis (when $u\ge 0$) in these coordinates, which is likely
the source of the discrepancy, and implying a well-behaved limit procedure
is needed to compute the correct ADM quantities (as indeed was the case for
the studies~\cite{1998CQGra..15.3841A,2000CQGra..17.3645A} mentioned above, where a
limit sequence based on a family of boosted Schwarzschild black hole spacetimes was used).

The equivalent expression for $^{(x)}\Psi_4$ (\ref{psi4_brinkmann}) is more complicated
in the coordinates (\ref{metric_gen}). We will not reproduce the full expression
here, though it is illuminating to consider the simpler case of
an axisymmetric wave, namely $h=0$ (\ref{efe_cyl}) and $g_m=0$ (\ref{g_decomp}):
\bea\label{xpsi4}
 ^{(x)}\Psi_{4,m=0}&&\\
=2\pi f(u)&&\left(\frac{2 q(\rho)}{\rho}-g_0(\rho)\right)
             \left[\cos(2\theta) + i \sin(2\theta) \right] \nonumber
\eea
where we have substituted in (\ref{q_eqn}). Note that the spatial tetrad with respect
to which we have defined $^{(x)}\Psi_4$ is still aligned with the $(x,y,z)$ coordinates,
as with (\ref{psi4_brinkmann}); hence, the $\theta$ dependence in (\ref{xpsi4}) is 
due to the rotation of this tetrad relative to the $(\rho,\theta)$ coordinate basis
(i.e. the metric distortion induced by this gravitational wave is axisymmetric by construction,
and would be manifestly so as measured by $\Psi_4$ built out of a tetrad 
aligned with the $(x,\rho,\theta)$ coordinates).
For an infinite, homogeneous matter
source with $g_0$ a constant $\alpha_0$, $q=\rho\alpha_0/2$ and, as before,
$^{(x)}\Psi_{4,m=0}$ vanishes and the spacetime has no Weyl curvature. However,
for a source with finite transverse extent, outside the region of matter
where $g_0=0$, $q\propto 1/\rho$ and so $^{(x)}\Psi_{4,m=0}\neq 0$. Thus, a localized,
plane-fronted null fluid wave does act as a source of ``pure'' plane-fronted gravitational
waves, it is just that the Weyl curvature happens to vanish within an inner core
about the axis if that core has constant matter density, and the matter distribution
remains axisymmetric throughout the spacetime.

\subsection{Fluid to particle distributions}\label{sec_dust}
Though a null fluid is useful to understand plane-fronted gravitational
wave spacetimes propagating in a single direction, this matter model is not 
ideal to explore collisions
of such waves. The reason is such matter easily
forms caustics, where the assumption that a single, unique velocity vector
field describing the fluid flow exists, breaks down. Correspondingly,
the Euler equations governing the flow break down at the caustic, and the solution
cannot be uniquely extended beyond the caustic. This problem
of the lack of a unique velocity vector is actually more pronounced than merely
being associated with caustics, as can be seen imagining the case where we
collide two identical distributions head on, as follows. A pressureless fluid
does not self-interact, and in the limit of weak gravity where the fluid
is propagating in flat space, we thus expect the two opposing streams
to simply pass through each other. This implies that, in the lab frame
where both streams are observed to have identical energy profiles,
there will be a moment of time symmetry as they pass through each
other where instantaneously the superposed
profiles have zero velocity. This is impossible to realize using
a single null fluid with stress-energy tensor of the 
form (\ref{efe}) (i.e., a null vector is incapable of describing a moment of time symmetry).
The problem can be remedied for this particular scenario by considering
two independent, noninteracting fluids, one describing the right propagating
pulse, the other the left. The stress-energy tensor is a sum of the two
individual fluid stress-energy tensors, and this sum will accurately
reflect the moment of time symmetry. In particular there will be no momentum
density, but there will be anisotropic pressure (pressure along the collision axis,
none transverse to it)\footnote{The situation is somewhat different
if a finite-$\gamma$ collision is considered with timelike fluid stars.
If the two stars are modeled with separate ideal fluids
that only interact gravitationally, at the moment of time symmetry in a collision
the net pressure tensor will also be anisotropic. However, here, 
a single fluid model can still be adequate to model the collision (as used in~\cite{East:2012mb}), 
since at the moment
of time symmetry the timelike vector describing the fluid flow will
simply coincide with the lab-frame's velocity 4-vector. Though in this latter
case the pressure tensor must still be isotropic. In the high speed
limit there will be a much steeper pressure gradient along
the collision axis (due to the length contraction of each star in the lab frame)
than transverse to it, which postcollision will cause evolution of the fluid
to proceed in a manner at least qualitatively consistent with the case
of two noninteracting fluids (though of course the two models will give very
different outcomes in the low velocity limit).}.
However, when including gravity, a two-fluid model
would only be a temporary fix, as the gravitational interaction between
the streams will cause focusing, eventually leading to caustics in each flow.

We therefore decide to treat the matter as a distribution of collisionless
particles instead, i.e. as governed by the collisionless Boltzmann (or Vlasov) equation.
We approximate this using a particle treatment,
where 
the energy distribution is given by the sum of a large 
number of particles, each locally traveling along a geodesic.
This easily resolves the uniqueness issues associated with a single, global fluid
vector field. 
However, this introduces a new problem of what exactly we mean by
a ``particle.'' 
In Newtonian gravity, the treatment of a pointlike distribution 
of energy is straightforward. In contrast, in general relativity, there is no simple
analogue: putting all the matter
at a point, however small the rest mass, produces a black hole.
That suggests one viable particle model
is a collection of black holes for timelike particles, and presumably
this could be extended to null particles by locally taking AS limits for
each black hole. However, a rigorous (without approximation) implementation
of this idea shifts all ``matter'' to ``geometry'' on the left hand side
of the Einstein equations, and certainly is not something practical to implement
numerically, even for one AS particle as discussed above, let alone a large enough 
number to approximate a continuum energy distribution.

The second, more practical option, is to treat each particle as if it were 
some form of solitonic matter, and then define some kind of averaging operation 
that consistently
adds the contribution of each particle to the discrete representation
of the stress-energy tensor used in the Einstein field equations.
(Or, alternatively, we can think of our particles as a discrete sampling
of the underlying continuum distribution described by the Boltzmann equation, 
and the problem is how to reconstruct the stress-energy tensor from this sampling.)
The easiest way to do this would be to assume the particle's
characteristic radius is much smaller than any mesh cell we will use in a numerical scheme,
and the particle's energy is sufficiently small that any self-force effects 
are negligible compared to numerical truncation error.
In principle, one could consider larger particles, as in Newtonian
smooth particle hydrodynamic codes. However, finite size effects might
then need to be considered. Moreover, for particles moving relativistically
length contraction must be taken into account, which complicates
the averaging operation for particles that span multiple cells.

We finish this section by writing the relevant definitions
and equations for the particle model, 
with the averaging for the stress tensor implied, but not explicitly stated;
in Sec.~\ref{sec_calc_set} we describe the particular averaging procedure we use in the code.
The stress-energy tensor for a collection of $N$ particles schematically takes the form
\be\label{set_dust}
T^{ab}=\sum_{i=1}^N\epsilon_i \ell_i^a \ell_i^b,
\ee
where $i$ labels the $i^{th}$ particle, traveling along a geodesic curve $x_i^a(\lambda)$ with
$\lambda$ the affine parameter along the curve, and $\ell_i^a$ the tangent to the curve:
\be\label{null_def_dust}
\ell_i^a \equiv \frac{d x_i^a(\lambda)}{d\lambda}, \ \ \ \ell_i^a \ell_{i a}=-m_i^2,
\ee
(no summation over particle label $i$). 
Here $\epsilon_i$ is a function that is related to the energy density
of the particle, the choice of affine parameter, and also the averaging process.
For null particles, the rest mass $m_i$ is zero, while for the massive case, the affine
parameter is chosen such that the proper time $\tau=m_i\lambda$. 
Each particle follows a geodesic
of the spacetime
\be\label{geod_eqn}
\ell_i^b \nabla_b \ell_i^a=0.
\ee
In an appropriate limit of an infinite number of particles, this 
model should reduce to a collisionless Boltzmann model. Of course,
directly discretizing the position-momentum phase space and evolving the
density function
would also be a viable approach to study the ultrarelativistic scattering
problem. The computational difficulty to solving the Boltzmann equation
is the additional dimensions required to represent the distribution in phase space.
On the flip side, the computational shortcoming of a particle
model is the slow $\sqrt{N}$ convergence to the desired continuum limit.

For distributions that can be described by a single-valued velocity vector field,
the particle model is equivalent to the fluid model discussed above for single wave fronts.
In the code, as described in the following section, we use the fluid
model to provide initial data, then evolve using the particle model. We always
use a sufficient number of particles in the latter so that
the large $N$ approximation to the continuum limit is smaller than
numerical truncation error from the discretization of spacetime, 
as judged by convergence of the constraint equations.

\section{Numerical Code}\label{sec_code}

In this section we describe the numerical code, reviewing
the generalized harmonic formalism we use for representing
the Einstein equations (Sec.~\ref{sec_metric_evo}); some aspects
of initial data (Sec.\ref{sec_id_gen}) and gauge
choice (Sec.~\ref{sec_gauge_gen}), leaving details
to Sec.~\ref{sec_results}; how we integrate geodesics (Sec.~\ref{sec_geod_evo});
and how we compute the stress-energy tensor from a distribution
of geodesics (Sec.~\ref{sec_calc_set}).
We focus on the case of colliding null waves, but besides the 
discussion of initial data and gauge conditions, the code is generally
applicable to both null and timelike (massive) particles, and we demonstrate
an application to an inhomogeneous matter-filled universe in the appendix.

\subsection{Metric evolution}\label{sec_metric_evo}
For the metric tensor, we solve the Einstein equations using the generalized harmonic formalism:
\bea\label{efe_harm}
g^{cd} g_{ab,cd} &+& \nonumber \\
g^{cd}{}_{,(a} g_{b)c,d} +2 H_{(a,b)} &-& 2 H_d \Gamma^d_{ab} + 2 \Gamma^c_{db} \Gamma^d_{ca}\nonumber\\
&=& - 8\pi (2 T_{ab} - g_{ab} T),
\eea
where $\Gamma^a_{bc}$ is the metric connection, $T$ the trace of the stress-energy
tensor, and $H^a$ are the so-called source functions, defined
by 
\be\label{harm_def}
H^a \equiv \Box x^a.
\ee
Here round brackets denote symmetrization and $\Box\equiv \nabla_a \nabla^a$.
During evolution, the source functions are treated as independent
functions, and can be thought of as encoding the coordinate freedom 
of the spacetime. Therefore, additional conditions/evolution equations need
to be supplied for them, with the definition (\ref{harm_def}) then becoming
a constraint 
\be\label{harm_const}
C^a\equiv H^a - \Box x^a = 0 \ .
\ee
In fact, the time derivative of this constraint essentially gives the usual
Hamiltonian and momentum constraints (see, for
example,~\cite{Lindblom:2005qh}).
We numerically solve (\ref{efe_harm}) with constraint damping terms~\cite{Gundlach:2005eh} (together with the gauge evolution equations
discussed later) 
utilizing a $4^{th}$ order accurate finite difference code with adaptive mesh refinement (AMR),
using the methods and techniques described in~\cite{Pretorius:2004jg,Pretorius:2005gq,code_paper}.
(Though since we presently only employ a second order accurate calculation of the stress-energy
tensor from the particle distribution, as described below, the overall accuracy of the
code is second order in the continuum limit.)
We use Cartesian-like coordinates where $x^a=(t,x,y,z)$, and for the axisymmetric evolutions
presented here use the modified Cartoon method~\cite{Alcubierre:1999ab} introduced
in~\cite{Pretorius:2004jg}. The modified Cartoon approach still uses the Cartesian form of the metric,
though only evolves a single $\theta={\rm constant}$ slice of the spacetime.

One difference for these simulations compared to earlier studies performed with this code, 
is here we do {\em not} spatially compactify the coordinates. The
reason is that with our initial data (see Sec.~\ref{id_specific} for the explicit solutions)
the asymptotic form of the metric, though asymptotically Minkowski, is not in the
usual trivial Cartesian form of $ds^2=-dt^2 + dx^2 + dy^2 + dz^2$.
Compactification transforms the metric into a representation that is singular
at infinity, but beginning from this form of the metric, the singular portion is easy to 
factor out analytically, with the code then only storing the regular part. 
Though this would, in principle,
be possible to do with the null wave initial data as well, given the nontrivial,
initial-data-dependent structure of the metric at infinity, it would have required
significant updates to the code. The disadvantage to not compactifying is then specifying
consistent, physically correct outer boundary conditions becomes challenging. 
We bypass this issue by placing the outer boundaries in the uncompactified code
sufficiently far away that they are out of causal contact with the inner region
of the domain where we will measure properties of the solution.

\subsection{Initial data}\label{sec_id_gen}
\begin{figure}
\vspace{0.1in}
\includegraphics[width=3.50in,draft=false,clip]{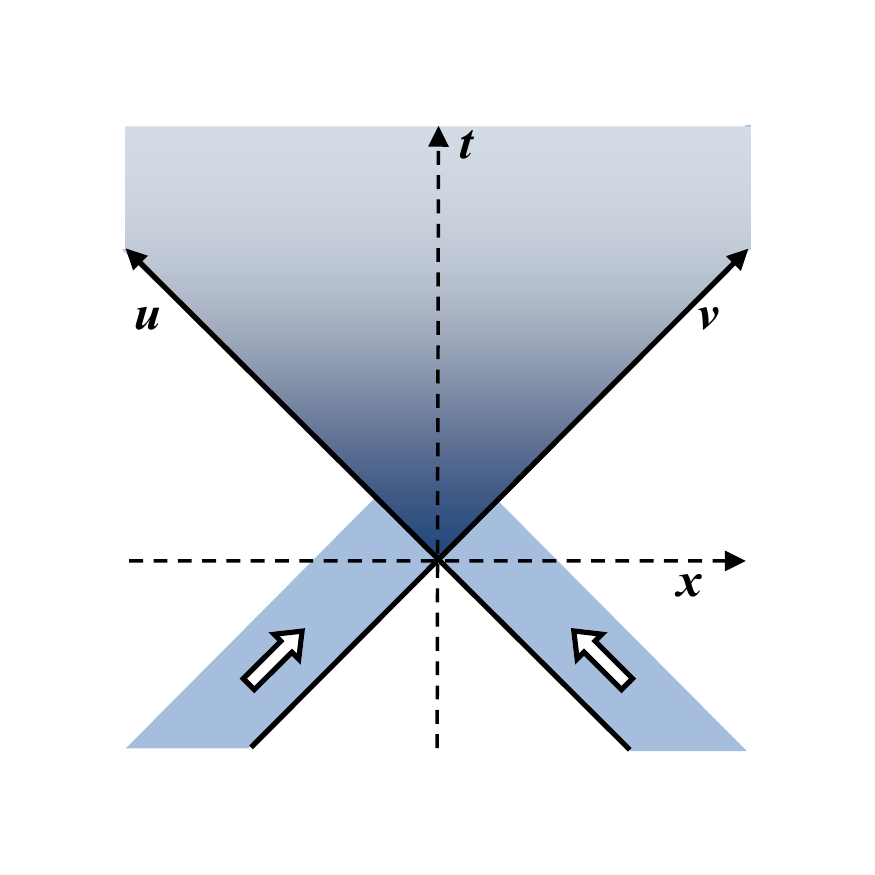}
\caption{
A spacetime diagram depicting the collision of 
left and right moving null-fronted plane waves.
The nonlinear interaction occurs in the
region $(u>0,v>0)$, depicted by the darker shaded region.
To the past of this, the spacetime geometry is either that of one of the
null waves (lighter shaded regions), or Minkowski spacetime.
}
\label{fig_2_pulse}
\end{figure}

For initial data, we superpose two null plane-fronted waves, one propagating
to the right (the $+x$ direction), the other to the left (the $-x$ direction); 
see Fig.~\ref{fig_2_pulse}.
The right (left) moving wave has compact support in $x<0$ ($x>0$) at $t=0$
(i.e. they do not overlap), and each is Minkowski spacetime on 
either side of the pulse.
For the pulse moving to the right we use the form of 
the metric and stress tensor discussed in Sec.~\ref{reg_plane_waves},
transforming to Cartesian coordinates using\footnote{As described in 
Sec.~\ref{id_specific}, for the runs presented here
we further transform the metric far to the left (right) of the right (left)
moving pulse to alleviate some of the resolution issues that might otherwise arise,
but that is immaterial to the discussion here.}
\be\label{to_cart}
u=t-x,\ v=t+x,\ y=\rho\sin\theta, z=\rho\cos\theta.
\ee
The boundary conditions for the integral defining $\beta(u)$ (\ref{beta_eqn})
are chosen so that the metric to the right of the pulse is in 
standard Minkowski form $ds^2=\eta_{ab} dx^a dx^b\equiv-dt^2 + dx^2 + dy^2 + dz^2$
(as discussed in Sec.~\ref{reg_plane_waves}, the form of the metric
which is $\eta_{ab}$ on {\em both} sides of the pulse has a $\ln\rho$
divergence within the plane of the pulse).
For the pulse
moving to the left, an analogous solution is used, but with the nontrivial
metric and matter functions depending on $v$ instead of $u$, and boundary
conditions for the analogous $\beta(v)$ flipped so that
the metric is $\eta_{ab}$ to the left of the pulse. Then,
consistent initial data for the Cauchy evolution performed by the code
is trivial : at $t=0$ the solution for $x\le0$ is exactly that of the right
moving pulse, and for $x\ge0$ is exactly that of the left moving pulse.

\subsection{Gauge conditions}\label{sec_gauge_gen}
For the source functions that define the gauge in the generalized harmonic formalism,
we begin with the gauge of the initial, superposed exact plane-fronted wave solutions. 
This superposed gauge is not adequate to use after the interaction of the 
waves, and so, within an inner volume of the domain where the interaction takes
place, we smoothly transition to a variant of the damped harmonic gauge used
in earlier high speed soliton collision simulations~\cite{Choptuik:2009ww,East:2012mb}.
The explicit form of the initial gauge source functions and source function
evolution equations are given in Sec.~\ref{sec_gauge}.

\subsection{Geodesic evolution}\label{sec_geod_evo}
We solve the geodesic equation~(\ref{geod_eqn}) for each particle as follows
(dropping the particle label $i$ here for clarity).
Instead of evolving the position $x^a$ as a function of affine parameter $\lambda$, we directly
integrate as a function of coordinate time $t$, using $dt/d\lambda=\ell^t$.
Then the geodesic equation can be reduced to the following system of
seven first order ordinary differential equations (ODEs), where we use the overdot ($\dot{\ }$) to denote $\partial_t$:
\bea
\dot{x}^j = \ell^j/\ell^t\nonumber \\
\dot{\ell}^a = - \Gamma^a_{bc} \ell^b \ell^c/\ell^t
\eea
where the metric connection $\Gamma^a_{bc}$ is evaluated at the 
location of the geodesic $x^a$.
In the code, we integrate these equations using a $4^{th}$ order
Runge-Kutta scheme. We have the option to 
enforce the normalization condition $\ell^a \ell_a=-m^2$ after each time step; there is no
unique way to do this, and we have chosen to solve
the normalization constraint for $\ell^t$. If the constraint is not explicitly
enforced during evolution, then we can use it as a diagnostic 
to check that it does converge to zero at the expected rate;
similarly if we do enforce it, we can use the correction
induced to $\ell^t$ after each time step as the diagnostic quantity
that should converge to zero. In tests we have conducted, both schemes
perform similarly over the relatively short time scale geodesic evolutions presented
here (which for most geodesics is much shorter than the net run time
as they are removed from the domain when they enter the excised region
inside the apparent horizon), though constrained evolution seems
to produce more accurate results at fixed resolution for longer time evolutions.

Initial conditions $x^a(t=0)$ and $\ell^a(t=0)$
are chosen so that the initial distribution of particles
produces a stress-energy tensor (as discussed in the following section)
giving a consistent sampling of the desired fluid continuum limit~(\ref{cont_set}).

\subsection{Calculating the stress-energy tensor}\label{sec_calc_set}
There are two issues that we need to address when calculating the effective
stress-energy tensor used in the Einstein equations coming from the distribution of particles.
First, how to add the contribution of a particle to the stress-energy tensor of
the cell containing it. Second, how to efficiently incorporate
this averaging process into the Berger and Oliger (BO) AMR algorithm we use, specifically
as it relates to time subcycling, which naively would seem to require
integrating the same geodesic multiple times on all resolution levels it 
overlaps (as the BO algorithm does for continuum evolution equations).

\subsubsection{Averaging}
The averaging procedure is 
how we convert the stress-energy tensor of a single particle 
into an equivalent cell-based representation such that 
we will obtain the same solution to the Einstein equations 
in the continuum limit as adding the contributions from a smooth distribution
of particles.
This can be quite complicated if we need high order accuracy, which nominally
would entail distributing a finite-sized model of the particle smoothly over 
a set of cells, taking 
the variation in the spacetime geometry into account.
For these initial studies, we are not concerned with high
order convergence; second order will suffice, and so we can avoid 
all these complications by simply smoothing a particle to
a single containing cell. 

To implement this, we demand
that the stress-energy tensor at some moment of coordinate
time $t$, integrated over the proper
volume $\Delta V_p$ of the cell containing the geodesic,
gives the same energy/momentum that an observer in the 
reference frame of the 
simulation would measure the particle to have. Here, the relevant observer is
that traveling along the unit timelike vector $n^a$ normal
to $t={\rm constant}$ surfaces, and the corresponding proper
volume element is $\Delta V_p=\sqrt{h} \Delta V$, where $h$ is the
determinant of the spatial metric (not to be confused with the metric
function $h(\rho,\theta)$ used earlier in the discussion of plane wave solutions), 
and $\Delta V=\Delta x \Delta y \Delta z$
is the coordinate volume element. A straightforward calculation
shows that the contribution of a single particle 
with 4-momentum $\ell^a$ 
to the effective continuum stress-energy tensor ${}^{(c)}T_{ab}$ of the cell
containing it is
\be\label{set_average}
{}^{(c)}T_{ab} = \frac{1}{\sqrt{h} \Delta V} \frac{\ell_a \ell_b}{(-\ell_d n^d)} \ .
\ee
To check this, recall that an observer with 4-vector $n^a$ measures the energy density 
of the stress tensor to be $T_{ab} n^a n^b$, and measures the total energy
of the particle to be $-n^a \ell_a$.
The net continuum stress-energy tensor is then just obtained by
summing the contribution from all particles in the cell.
Referring back to the schematic form of the particle stress tensor
written in Eq. (\ref{set_dust}), the equation above defines exactly
what we mean by $\epsilon_i$. Note that it is not a constant, 
and its particular value is affected by our choice of affine parameter,
defined to let us interpret $\ell^a_i$ as the physical, 4-momentum
of the particle. For a 
set of timelike particles, again, $\ell_i^a=m_i u_i^a$ is the 4-momentum
of a particle of rest mass $m_i$ and 4-velocity $u_i^a$.

In our code, we discretize the Einstein equations using a vertex-centered
mesh. To obtain a second order accurate representation of the vertex-centered 
stress tensor, we take the above computed ${}^{(c)}T_{ab}$ for
each particle, and distribute it to the surrounding vertices, linearly
weighting the contribution to each vertex based on the distance
of the geodesic from it.

\subsubsection{An efficient evolution scheme within the Berger and Oliger time subcycling algorithm}

The BO AMR algorithm~\cite{Berger:1984zza} uses a grid
hierarchy consisting of nested grids, where finer resolution (child) grids are entirely
contained within coarser resolution (parent) grids. A collection of grids
at the same resolution is called a {\em level}. Hyperbolic differential
equations discretized on such a hierarchy are evolved in time using
the following algorithm (for a more detailed description and pseudocode
see~\cite{Pretorius:2005ua}, for example). For simplicity, assume the refinement
ratio between successive levels is two, and the same Courant-Fredrichs-Lewy (CFL)
factor is used on all levels (as used in our code). Beginning from the same
starting time, on any given level, one time step of size $\Delta t$ is taken on all the grids 
at that level {\em before} two times steps of size $\Delta t/2$ are taken on the grids of the 
child level.
This rule is applied recursively. 

The reason for this scheme is so that the solution from
the parent level can be used to set boundary conditions at child level interior
boundaries, via interpolation in time from the parent solution. [At computational domain
boundaries the relevant partial differential equation (PDE) boundary conditions are 
always applied.] Also,
this time subcycling is optimally efficient for hyperbolic PDEs in the sense
that each level is evolved with the same CFL factor (as opposed to 
schemes where all levels, or a single level but with nonuniform grid cells,
are evolved with the same time step; the effective CFL
factor for coarse levels/cells in such an approach could be much smaller than necessary
for stable evolution).
Note also that on parent levels the unigrid evolution is applied everyone, even at regions
where a higher resolution child grid is available; the subsequent solution from the child grid
evolution is injected back into the parent level after they both are in sync again,
so the discrete solution after each global time step is always single-valued. This might
seem like a waste of computational resources, though it is a relatively minor
expense (most of the computation happens on the finest levels covering any region),
and significantly simplifies development of AMR capable codes in that the
underlying PDE evolution scheme can be almost completely ignorant of the mesh
hierarchy (it only needs to know which boundaries are physical vs interior, 
and for interior boundaries it simply leaves the corresponding points untouched).

We would now like to include our geodesic integration within the BO AMR algorithm,
and following the spirit of the algorithm, we want to solve the coupled
Einstein-particle equations on each grid in a manner which relies minimally on knowledge 
of what mesh hierarchy the grid is part of. However, the problem here is
because the geodesics are lines through the spacetime, and are integrated via ODEs, 
they do not have the same natural multiscale representation on a BO mesh hierarchy
that continuum functions, such as the metric or stress tensor, have. Yet, we still
want the solution to the geodesics to be computed using metric values from the finest
mesh containing the geodesic. 
How then do we evolve geodesics, and use them to define the stress-energy tensor,
within the recursive, time subcycling
algorithm? One simple option is just restart each geodesic for every finer
level containing it, with the last then giving the most accurate solution that is kept.
The problem with this is it will be very computationally inefficient for a deep
hierarchy, as
we do not have a multiscale representation of the set of geodesics; i.e.
each geodesic is integrated $L$ times if the depth of the hierarchy
is $L$ at the location of the geodesic. (This problem is mitigated
for mesh-based PDE evolution because of the multiscale representation---effectively
the number of mesh points where the PDEs are multiply integrated
drops as $1/2^d$ per level if the refinement ratio is $2$, and there are $d$ spatial dimensions.)

One workaround would be to introduce a multiscale sampling of the
continuum matter distribution where, say, the geodesic number density per cell
is kept fixed going from level to level, and a coarse-level geodesic
is some average of $2^d$ fine-level geodesics. This would
complicated the structure of the AMR driver significantly.

The solution that we take instead is to adapt the scheme proposed
in~\cite{Pretorius:2005ua} for evolving a system of elliptic-hyperbolic
PDEs within the BO time stepping framework. A similar problem to the
above arises for the elliptic equations, and without going into detail
here, the solution is to {\em not} solve the elliptic equations when 
{\em descending} the recursive tree (going from course to fine), when the hyperbolic
equations are solved. Rather, then values for the elliptic variables
are extrapolated from prior time levels, and instead the elliptics are solved
when {\em ascending} the recursive tree (when hyperbolic
variables are injected from fine to coarse levels). The way this algorithm
is adapted to particles is when descending the recursive tree and the Einstein
equations are solved, the stress-energy tensor used on the right hand side in each cell 
is either (a) extrapolated from
past time levels in regions where finer levels exists, and in these cells
no geodesics are integrated, or (b) the geodesics within the cell are integrated 
(this is thus the finest level containing them) and used to compute the stress
energy tensor for the neighboring vertices as described above. When ascending
the tree, the most accurately available stress-energy tensor values are
injected back up to coarser levels.

The reason this approach was simpler for us is the AMR driver
code we use (PAMR/AMRD code~\cite{pamr_amrd}) already has infrastructure to
handle the extrapolation: we simply define the stress-energy tensor as 
if it were an elliptic variable in the code, and all the unigrid geodesic
integration code needs to know is whether a given cell is the finest
cell: if so, the geodesics within it are integrated and the stress tensor
computed there. In the present code we save two past time levels,
which is adequate to allow second order accurate extrapolation, and maintain
overall second order accuracy of the evolution (the metric and geodesics are
still evolved with $4^{th}$ order accurate Runge-Kutta, but the stress-energy
energy calculation discussed above is only $2^{nd}$ order accurate, independent
of the way we evolve the geodesics within the BO framework).

In the remainder of the main text, we will focus on an axisymmetric, null
particle case. However, we have also tested this code for 3D (i.e.
nonaxisymmetric) calculations, as well as with massive (timelike) particles,
as we demonstrate in the appendix with results from a simple inhomogeneous
cosmology setup.

\section{Results: Black Hole Formation in axisymmetric, head-on collisions}\label{sec_results}
Here we present results from simulations of the head-on collision of
two axisymmetric null dust-sourced plane gravitational waves.
We select initial data to closely link to the Aichelburg-Sexl limit.
As discussed in the introduction, Penrose first looked at the AS case.
Assuming a mass based on the initial apparent horizon area is a lower limit to the
mass the black hole settles down to, and hence the difference between the initial
spacetime and apparent horizon mass is an upper limit to the gravitational radiation
$E_{\rm GW}$ emitted during the collision, Penrose's computation
gives $E_{\rm GW}<29\%$.
D'Eath~\cite{DEath:1976bbo}, and D'Eath and Payne~\cite{1992PhRvD..46..694D}
perturbatively explored the far-field region of the AS limit,
and were able to directly estimate the gravitational wave emission,
arriving at $E_{\rm GW}=16.4\%$. Approaching the AS collision from finite-$\gamma$
timelike compact object collisions, \cite{Sperhake:2008ga} first collided black holes
up to $\gamma=2.9$, and extrapolating the results to $\gamma=\infty$, found
$E_{\rm GW}=14\pm3\%$. This scenario was extended to $\gamma\approx 7$ in ~\cite{Healy:2015mla},
where they estimated $E_{\rm GW}=13\pm1\%$\footnote{This might seem in mild tension
with our quoted number, though looking at Figs. 9 and 10 of ~\cite{Healy:2015mla}
suggests there is an effective systematic uncertainty associated with
the different classes of initial data and codes they use that might warrant a slightly more
conservative error estimate, and so we think there is not yet any significant indication that the
two different approaches will not reach the same AS collision spacetime in their
respective limits.}.
In \cite{East:2012mb}, compact fluid stars with up to $\gamma=12$ 
were collided, and it was found that $E_{\rm GW}=16\pm2\%$.

Similar to Penrose's first investigation where he found an apparent 
horizon at the moment of impact, earlier studies of null-radiation (gyraton)
interactions were able to show apparent horizon formation~\cite{Yoshino:2007ph}.
(These authors considered more general models of gyratons including rotation~\cite{Frolov:2005in};
the null-particle case maps to nonrotating gyratons.)
Here we are able to follow the evolution of spacetime through the formation of an apparent
horizon which eventually settles down toward a Schwarzschild black hole,
together with gravitational radiation streaming away from the collision.
Regarding the net gravitational wave energy emitted, as explained
below, we are not able to characterize the initial decay of the waves 
close to the collision (where we can measure them) in a manner that allows
extrapolation to infinity in order to accurately calculate the energy they contain.
Instead then, we assume $E_{\rm GW}$ is the difference between the initial
spacetime mass and late-time apparent horizon mass, the latter which we
can compute accurately, and using a conservative upper bound
of $0.1\%$ of energy in particles that escaped entrapment by the black hole
as an additional source of uncertainty, obtain $E_{\rm GW}=14.9\pm 0.8\%$.
As discussed more below, we expect this scenario to give 
$E_{\rm GW}$ that is a little below the AS limit by $\sim 0.3\%$.

The structure of the remainder of this section is as follows. In Sec.~\ref{id_specific}
we describe the specific initial data we use, in Sec.~\ref{sec_gauge} we discuss
our gauge source function evolution equations, and in Sec.~\ref{sec_case_study}
we present the results from the simulations, including convergence tests.

\subsection{Initial data}\label{id_specific}
Here we give the particular form of a single, right ($+x$) propagating null fluid
wave front we use for initial data in the numerical evolution (the left propagating
front has identical form, with $x\rightarrow-x$, as discussed in Sec.~\ref{sec_id_gen}).
For simplicity, here we show the pulse centered at $x=0$, though it is trivial
to shift it to any desired starting location at $t=0$.
We begin with the metric in the form (\ref{metric_gen}), and, since we are
restricting to axisymmetric spacetimes here, $h=0$:
\be
ds^2=-du dv + 2\beta(u)q(\rho) du d\rho + d\rho^2 + \rho^2 d\theta^2 \ .\label{metric_axi}
\ee
For the matter profile, we use a piecewise polynomial function in $u$, and a Gaussian
in $\rho$. 
Defining $\bar{u}\equiv u/\Delta u$, $\bar{\rho}\equiv \rho/\Delta \rho$, with
$\Delta u$ and $\Delta \rho$ constant parameters that define the scale of the 
profile,
\bea
\rho_e(u,\rho)&=& f(u) g_0 (\rho), \label{rhoe_def} \\
f(u)&=&\left\{\begin{array}{lc}
      0, & \bar{u}<-1 \\
      (\bar{u}^2-1)^2, & -1\leq \bar{u} \leq 1 \\ 
      0, & \bar{u}>1
     \end{array}\right.\\
g_0(\rho)&=&A e^{-\bar{\rho}^2}\label{g0_prof},
\eea
where $A$ is a parameter controlling the amplitude of the wave (note here that the $u$-extent of the
pulse is $2\Delta u$, a factor of 2 different from the discussion in Sec.~\ref{plane_waves}).
Equations (\ref{beta_eqn}) and (\ref{q_eqn}) then give 
\bea
\frac{\beta(u)}{4\pi\Delta u} &=&\left\{\begin{array}{lc}
      \bigg. 0, & \bar{u}<-1 \\
      \left[\frac{1}{5}\bar{u}^5-\frac{2}{3}\bar{u}^3+\bar{u}+\frac{8}{15}\right], & -1\leq \bar{u} \leq 1 \\ 
      \bigg. \frac{16}{15}, & \bar{u}>1
     \end{array}\right.\\
q(\rho)&=&A\frac{\Delta\rho}{2 \bar{\rho}}\left[1-e^{-\bar{\rho}^2}\right] \ .
\eea
The ADM mass for this spacetime evaluates to
\be
\bar{M} = A \frac{8\pi}{15} \Delta u \Delta \rho^2.
\ee
If we take
the point particle limit $\Delta u\rightarrow 0, \Delta \rho\rightarrow 0$, keeping
$\bar{M}$ fixed, (\ref{metric_axi}) becomes, for $\rho>0$,
\be
ds^2=-du dv + \frac{8\bar{M}}{\rho}\Theta(u) du d\rho + d\rho^2 + \rho^2 d\theta^2\label{metric_axi_as},
\ee
where $\Theta(u)$ is the Heaviside step function. For $\rho$=0, the metric in (\ref{metric_axi}) always
has $g_{u\rho}=0$ by regularity, but if instead we define (\ref{metric_axi_as}) as the metric including
$\rho=0$, then this is the AS solution if we identify its total energy $m$ with $\bar{M}$. 
Interestingly, as discussed in Sec.~\ref{reg_plane_waves}, if we were to take
(\ref{metric_axi_as}) (with $\bar{M}=m$) as a vacuum solution to the Einstein equations
and directly use it in the ADM formula (\ref{adm1}), the latter
would give an ADM mass of $m/2$. \footnote{Note that for a pure vacuum
case one could still use the metric ansatz (\ref{metric_gen}) with $h=0$ in axisymmetry, but would then
not impose the conditions (\ref{beta_eqn}) and (\ref{rhoe_eqn}) that otherwise seem to link
$\beta$ to $\rho_e$; instead $\beta(u)$ can then be considered the arbitrary function
we choose to specify the longitudinal extent of the pulse.}
For the matter sourced spacetimes with the above
energy profile, each line in the formula (\ref{adm1}) contributes $m/2$, but the
first line is identically zero for vacuum spacetimes. 

A further curious property of this solution is, transforming
to Cartesian coordinates via (\ref{to_cart}), the lapse function
is $\alpha = 1/\sqrt{1-\beta^2 q^2}$. The lapse becomes singular
when $\beta q=1$, implying the $t=\rm constant$ hypersurface
fails to be globally spacelike, and hence cannot be used
as initial data in a Cauchy evolution fixed to this time coordinate.
For the AS solution, this occurs to the left of the shock
for $\rho\leq 4 m$; for matter solutions this implies we
cannot use initial data in these coordinates if $\Delta \rho \lesssim 4 \bar{M}$.
Of course, this is just a coordinate singularity, though what
is curious about it is it seems to ``anticipate'' black hole
formation in a collision: colliding two identical
pulses each with mass $\bar{M}$, the total spacetime mass will be $2 \bar{M}$,
and the hoop conjecture argues a black hole should
form following collision if all the matter is focused inward within a hoop of
radius $4 \bar{M}$, the limiting estimate for $\Delta \rho$
for each pulse to have a well defined lapse.

An issue with the above coordinate system that relates to computational
cost is that to the left of the pulse there is nontrivial structure
in the metric within a region of size $\Delta \rho$ of the axis,
all the way to $x\rightarrow-\infty$.
Numerical experiments show that for long-term stable evolution this
region needs to be resolved with essentially
the same resolution as the inner part of the domain where the collision
occurs, even though the $x\rightarrow-\infty$ part of the spacetime
is simply Minkowski, with no dynamics. This is somewhat wasteful, and suggests
we should transform back to the $\eta_{ab}$ representation of Minkowski here. 
Though, as discussed in
Sec.~\ref{plane_waves}, we suspect one cannot transform to exactly this representation
after the wave while also maintaining $\eta_{ab}$ as the form of Minkowski ahead
of the wave (which is desired for the simplicity of initial 
data construction) and not introducing a $\ln\rho$ divergence as $\rho\rightarrow\infty$
within the plane of the matter. However, we can {\em partly} transform back to
$\eta_{ab}$ post-shock, in particular in a region
about the axis. In a sense, this can spread the nontrivial structure over a larger region
in $\rho$ that then needs much less resolution to resolve. To do this, before
transforming to Cartesian form from double null coordinates, we rescale $v$ following a
generalization of (\ref{ctrans1}):
\be
v=\bar{v}+G(\rho) L(u).
\ee
This transforms (\ref{metric_axi}) to
\be
ds^2=-du d\bar{v} + (2\beta q-G' L) du d\rho -G L' du^2 + d\rho^2 + \rho^2 d\theta^2\label{metric_axi_trans},
\ee
where, for simplicity, we do not show the functional arguments, and prime ($'$) denotes a derivative
with respect to the function's argument. To not affect the form of the solution
within the wave or ahead of it, we choose $L(u<u_{L1})=0$ for some $u_{L1}>\Delta u$,
and let $L$ smoothly increase to one over the region $u_{L1}<u<u_{L2}$ 
(in the numerical calculation we use a piece-wise $4^{th}$ order polynomial function for this).
For $G$ we choose $G'(\rho)=2\beta_0 q(\rho) T(\rho)$, where $G(0)=0$,
$\beta_0\equiv\beta(u=\infty)$, and $T(\rho)$ is a transition
function that is one for $\rho<\rho_{m1}$, and smoothly
(again via a piece-wise $4^{th}$ order polynomial in the numerical calculation) goes to zero
over the region $\rho_{m1}<\rho<\rho_{m2}$. The axis resolution issues
arose from the $du d\rho$ term in the metric; with this transformation we therefore eliminate
this term in the region $\rho<\rho_{m1}, u>u_{L2}$, and slowly reintroduce
it over a longer scale in $\rho$ controlled by $\rho_{m2}-\rho_{m1}$. The transformation
creates a new $du^2$ piece of the metric within $u=u_{L1}..u_{L2}$, though
its size can likewise be controlled by $u_{L2}-u_{L1}$.

\subsection{Gauge evolution}\label{sec_gauge}

At the initial time, and for a short time thereafter, 
the gauge source functions $H^a(t,x,y,z)$ are simply set
to the superposition of those of the exact solutions; 
i.e. (\ref{harm_def}) evaluated with the relevant version
of (\ref{metric_axi_trans}), after each is shifted in $u$ or $v$
to have the pulses at the desired starting positions, and then 
transformed to Cartesian coordinates. This simple gauge prescription
actually works even through collision for weak pulses that are
not close to forming black holes, but for strong pulses leads to 
a coordinate singularity some time after the interaction.
Therefore, at a specified time before the collision we
smoothly transition the gauge, over a chosen time period,
and within a chosen spatial volume of the origin, to
the damped harmonic gauge described in~\cite{Choptuik:2009ww} with
amplitude parameter $\xi$ (which is equivalent
to A15 in~\cite{Lindblom:2009tu} with $\xi=\mu_L=\mu_S$ and $p=1/2$). 

The evolution
is not particularly sensitive to exact parameter values, with $\xi \sim O(1/M)$ ($M$ is
the total mass of the spacetime),
and as long as the transition
in time takes place within $O(M)$ of the collision, and within a
volume that comfortably encloses the apparent horizon.  For the simulations
presented below, we use $\xi=2/M$, transition to damped harmonic within
$t=M/2$ (with a $4^{th}$ order piece-wise polynomial in time), and 
within a coordinate sphere of radius $r=150M$ of the origin. From $r=150M$ to
$r=200M$ the gauge smoothly (again with a $4^{th}$ order polynomial)
transitions to the superposed initial data gauge, and remains that from $r=200M$ to our
outer boundary at $|x|=y=250M$. The simulations are run until $t=175M$, and the two
pulses are set apart so that the interaction begins within $t\sim M/2$.

\subsection{Case study}\label{sec_case_study}
Here we present the results from simulations of the head-on collision
of two plane-fronted gravitational waves, sourced by null particles.
We choose initial conditions to connect to the shock-front AS collision
limit; i.e. large amplitude waves that promptly form a single encompassing black hole,
and whose characteristic width is smaller than its transverse length,
the latter corresponding roughly to the size of the eventual black hole. 
Specifically, each initial pulse is given a profile of the form (\ref{rhoe_def})-(\ref{g0_prof}),
with $\Delta u/\Delta\rho=1/8$, and $\Delta \rho = 3.125 \bar{M}$.
(This is close to the maximum compaction in $\rho$ we can choose in 
these coordinates; $\Delta \rho$ is a bit less than the
rough estimate of $4 \bar{M}$ given in Sec.~\ref{id_specific}, which would
be exact for a constant density with sharp $\rho$ cutoff.) The total mass of the spacetime
is $M=2 \bar{M}$. The facing edges of each pulse are 
set $\sim 1M$ apart, so that they begin to interact
within $t\sim M/2$, and an apparent horizon is first detected at $t=1.1M$.

Choosing large amplitude data with $\Delta u/\Delta\rho\ll 1$ results
in essentially all the matter falling into the black hole (some particles in
the Gaussian tail do escape), and so the postcollision spacetime
closely connects to the vacuum AS collision problem.
The true limit would take both $\Delta u$ and $\Delta\rho$ to zero, with
the matter distribution approaching a delta function. As discussion above,
we cannot make $\Delta\rho$ arbitrarily small with this class of initial data; however,
for $\Delta u/\Delta\rho\ll 1$, a black hole forms promptly,
and the gravitational interaction that happens outside the black hole
transverse to the collision is then, by causality, going to be independent
of how small $\Delta \rho$ is within the black hole (the asymptotic
falloff in $\rho$ outside the matter region is insensitive to $\Delta \rho$
for a sequence with fixed $M_{\rm ADM}$ and $\Delta u$). So in that sense,
in terms of $\Delta \rho$, we believe we are quite close to the AS limit.
Our initial data allows us to make $\Delta u$ arbitrarily small (for fixed
$M_{\rm ADM}$ and $\Delta \rho$). However, the smaller $\Delta u$ is,
the more computationally expensive the simulations become, as this
length scale needs to be resolved. In the limit $\Delta u\rightarrow 0$,
the geometry becomes shocklike, and this feature will persist (at least
along the leading edge of the shock) postcollision. In terms of
starting to resolve a shocklike feature, we are quite far from
the AS limit. (In a relative sense, any finite $\Delta u$ is always 
infinitely far from $\Delta u=0$; a more physical measure
might be the width of each initial wave $2\Delta u$ divided
by the final black hole diameter $\approx 4M$, which here equals $\sim 0.1$.)
Regarding net gravitational wave emission, we ran a 
preliminary survey decreasing
$\Delta u$ to as low as $\Delta u/\Delta\rho=1/32$, and extrapolating, this suggests
the  $\Delta u/\Delta\rho=1/8$ case underestimates the AS limit net energy emission by
$\sim 0.3\%$. However, these runs were performed with the same $h_0$ resolution (see below)
as the $\Delta u/\Delta\rho=1/8$ case presented in detail here, which effectively
means successively worse resolution for smaller $\Delta u$. Hence, the $\sim 0.3\%$ number
should be considered a rough estimate. Again, it would take
significant computational resources to explore
the AS limiting sequence in detail and accurately, and we leave that for future work.

Another practical reason for studying this point in parameter space
for this first application of the null particle code, is the black hole
that forms cleanly hides extreme focusing onto the axis 
with these axisymmetric profiles. The matter is pressureless 
(except for the effective anisotropic pressure that arises in multistream regions), and 
the initial data has no angular momentum about the symmetry axis, 
so even if no singularities form,
nothing prevents the distribution from focusing to very small
length scales, and to maintain convergence when this happens
requires computationally expensive, deep mesh hierarchies.

\begin{figure}
\vspace{0.1in}
\includegraphics[width=3.00in,draft=false,clip]{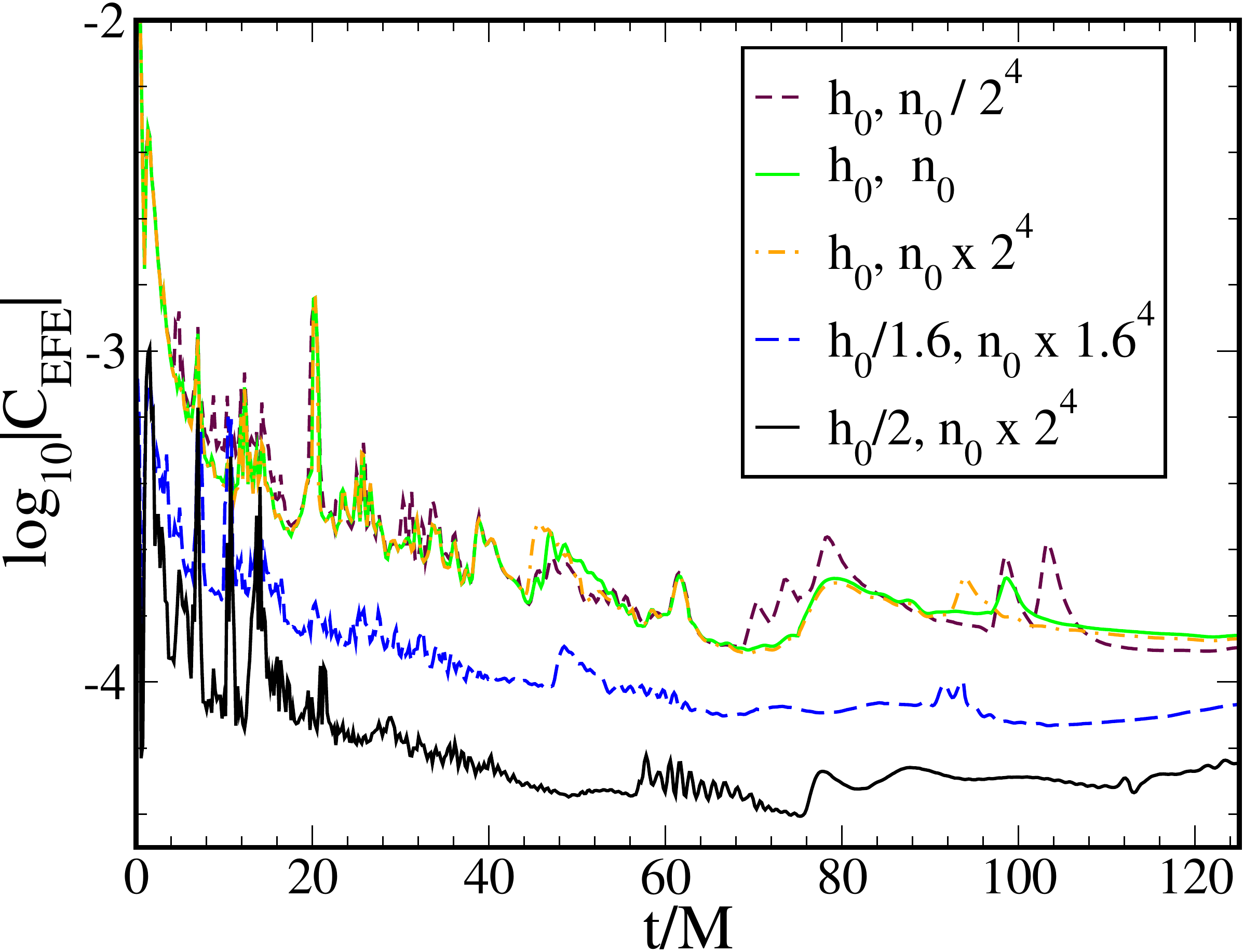}
\caption
{Log of the L2-norm of the constraints (\ref{harm_const}) over the inner
    $x\in[-125M,125M]$, $y\in[0,125M]$ portion of the computational domain
    (units/normalization arbitrary up to a constant shift).
Shown are three characteristic resolutions, with the
number of particles $n_0$ scaled as explained in the text when changing
the base resolution $h_0$. For the $h_0$ case, data from two
additional runs are shown with different numbers of particles,
demonstrating that here we are essentially in the domain
where grid-based truncation error is dominating the solution error.
The trends with resolution are broadly consistent
with $2^{nd}$ order convergence; 
some of the spikes in the higher resolution curves,
particularly noticeable for the highest resolution (solid black) curve at early
times, are due to the mesh refinement algorithm temporarily dropping a highest
resolution grid, and so during that time in the region about the black hole (which dominates
the constraint error)
the grid resolutions of the $h_0$ and $h_0/2$ simulations are actually the same.
Of course, we could have required the hierarchies to be identical
amongst the runs to give cleaner looking convergence plots, but as that is
not typically how we do runs, the above is a more representative example
of the convergence behavior of the code.
}
\label{fig_norm_efe}
\end{figure}

Our base case resolution (which we label with $h_0$) is such
that the coarsest level of the AMR hierarchy covers the entire
domain $x\in[-250M,250M]$, $y\in[0,250M]$ with cell widths of
size $1.5625M$, and has up to seven levels of 2:1 refinement,
giving a minimum possible cell width of $(1.5625/128)M$. The hierarchy
is dynamically generated via truncation error estimates. We evolve
until $t=175M$, and when measuring some property of the solution
restrict the measurement domain to be out of causal contact
with the boundary (assuming unit light speed), to mitigate
inconsistencies that may arise due to our outer boundary conditions
(which are given by superposed exact null wave solutions).
We sample each matter distribution with $n_0=3.84 \times 10^5$ geodesics,
which prior experimentation has shown is adequate to have the
$\sqrt{N}$ error be smaller than the discrete mesh truncation
error (and we also demonstrate that in Fig.~\ref{fig_norm_efe}).

\begin{figure}
\vspace{0.1in}
\includegraphics[width=3.00in,draft=false,clip]{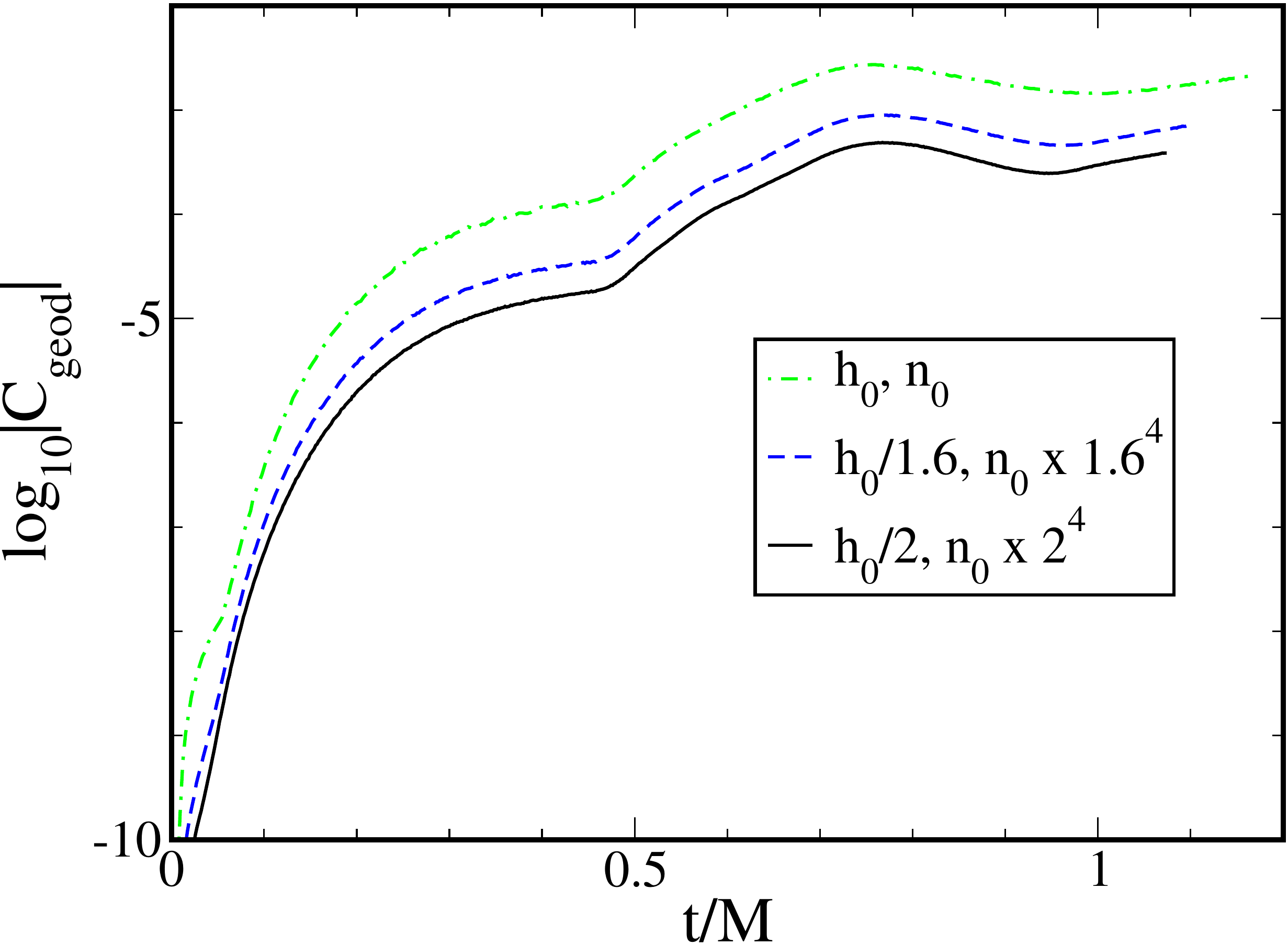}
\caption
{Log of the L2 norm over all geodesics of the normalization constraint $\ell_a \ell^a=0$ when 
performing a free evolution of the geodesic equation
(units/normalization arbitrary up to a constant shift).
The curves end once an apparent horizon is first detected, after which
we remove any geodesics (the vast majority of them for this
initial data) in the then excised region of the domain.
Constrained evolution gives lower norms for the corresponding
error calculated when adjusting the $\ell^t$ component of each geodesic to 
enforce the constraint, though for these short geodesic evolutions
the difference in the affect on the spacetime truncation error is negligible.
}
\label{fig_norm_geod}
\end{figure}

To check convergence---see Figs.~\ref{fig_norm_efe} and \ref{fig_norm_geod} for norms
of the Einstein and geodesic equation constraints respectively---and compute error 
estimates, we also ran simulations with 1.6 and $2\times$ finer
($h_0/1.6$ and $h_0/2$) base level resolution, adjusting the truncation
error threshold in the AMR algorithm to generate finer levels according
to the expectation of overall $2^{nd}$ order convergence. We also
in tandem change the number of particles by $n_0 \times l^4$ when
the spatial resolution is scaled by $h_0/l$. One factor of $l^2$ keeps
the particle density, hence $\sqrt{N}$ error the same in each cell, the additional factor
of $l^2$ is then to further increase/decrease the resolution of the sampling
within the cell to match the scaling of the mesh-based truncation
error. This highlights how computationally expensive a particle-based code is
to achieve high accuracy, and it is not clear in this respect
that it offers any advantage over directly discretizing the Boltzmann
equation in phase space (our main reason for going the former route is simplicity
of implementation). 

\begin{figure*}
\begin{center}
\includegraphics[width=3.00in,draft=false,clip]{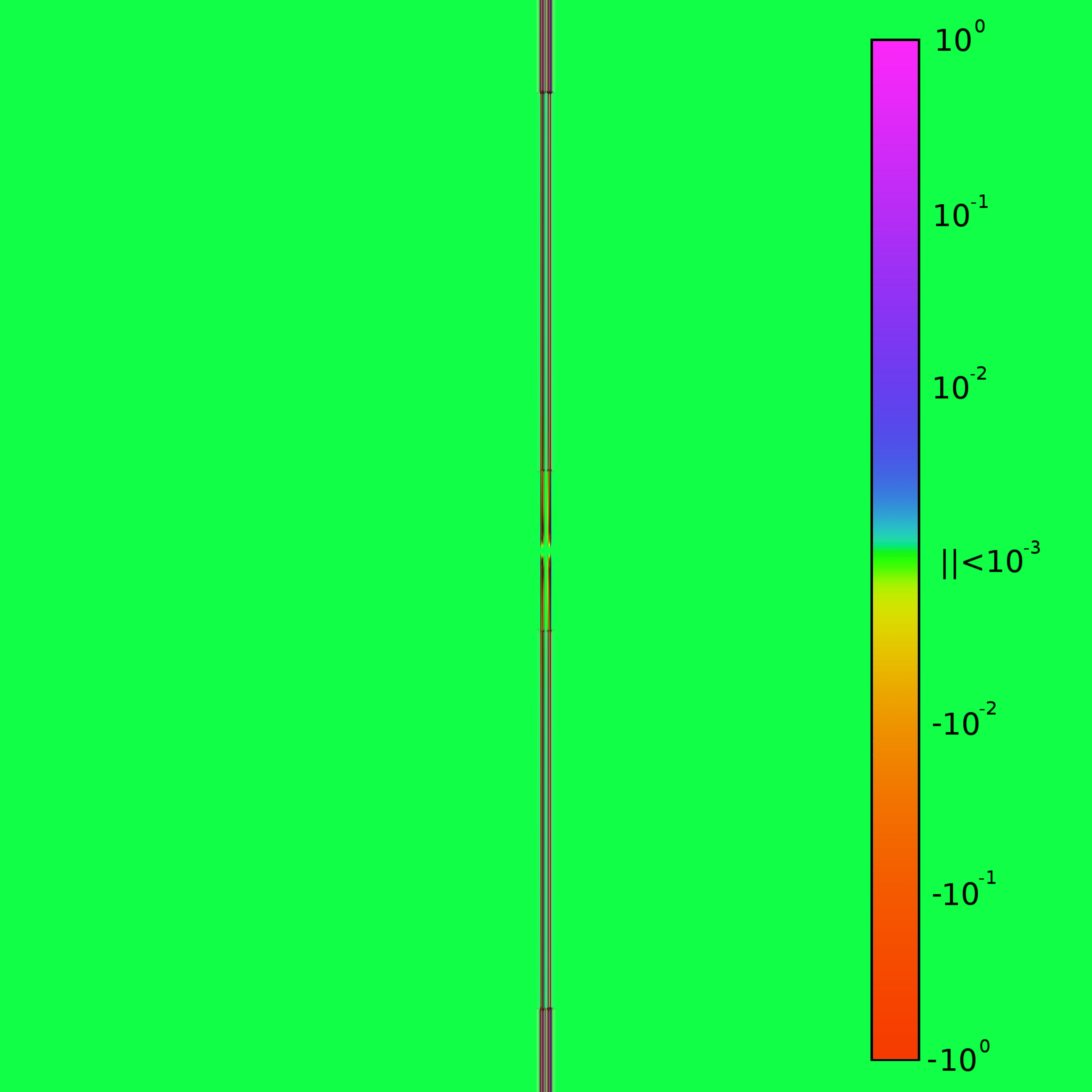}
\includegraphics[width=3.00in,draft=false,clip]{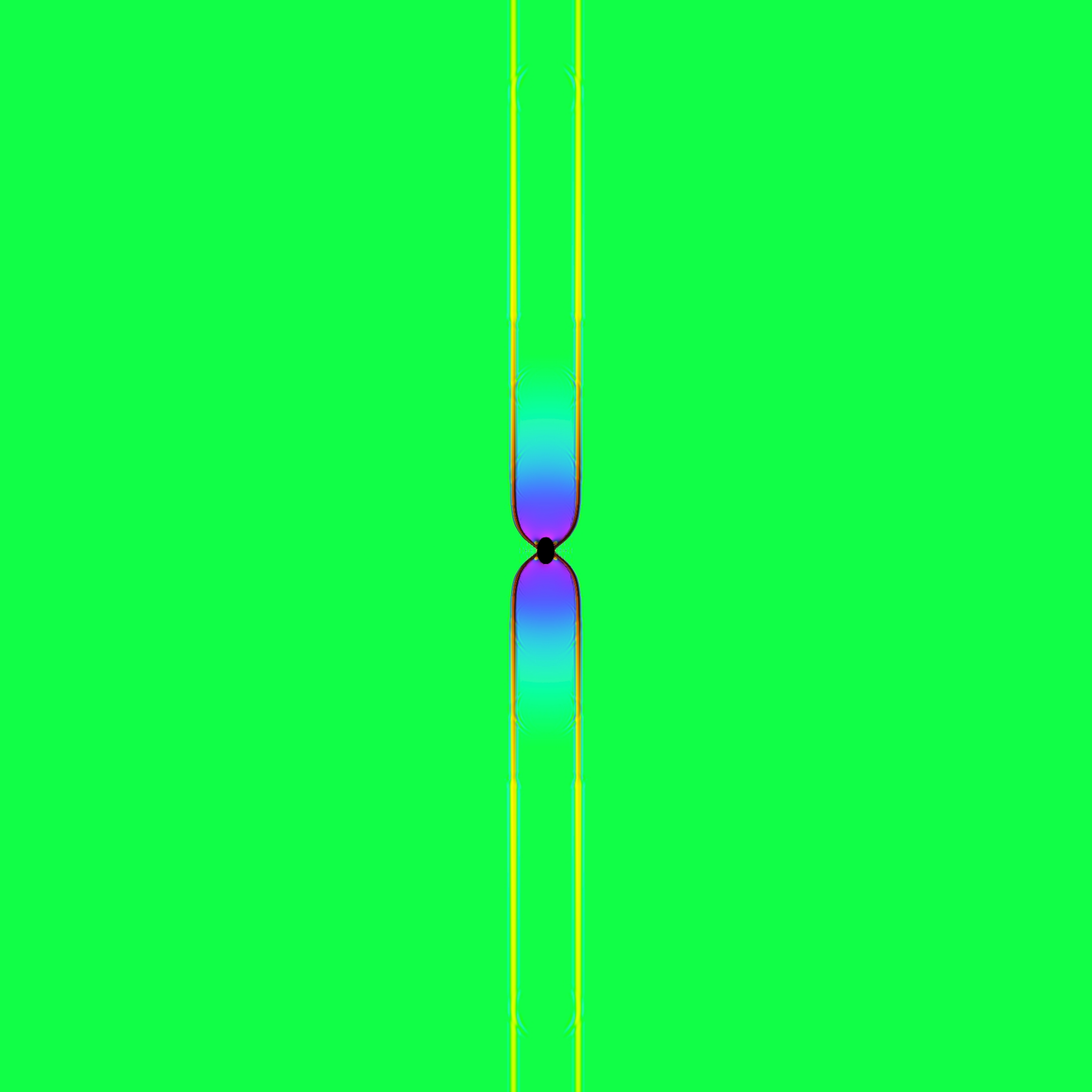} 
\\
\vspace{0.03in}
\includegraphics[width=3.00in,draft=false,clip]{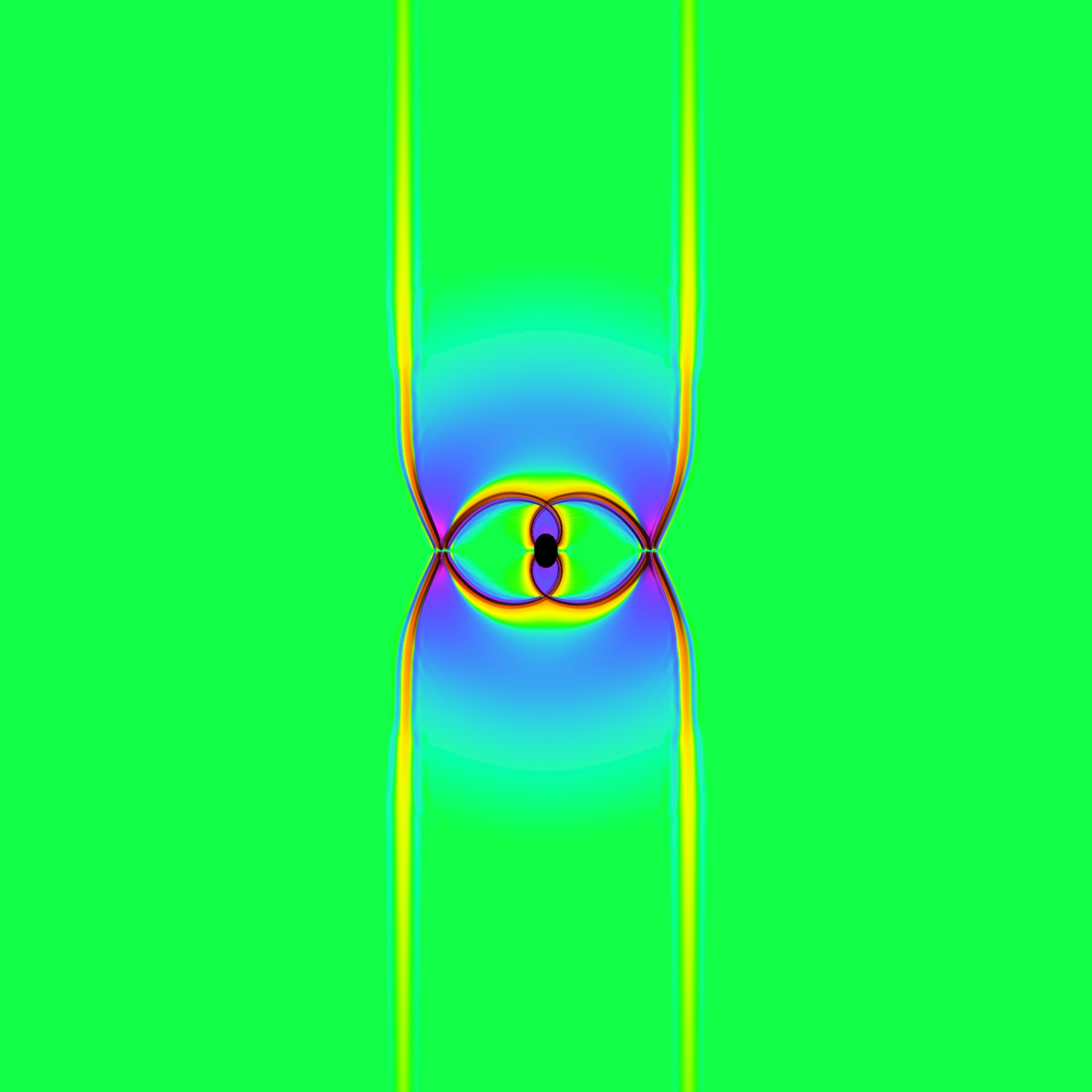}
\includegraphics[width=3.00in,draft=false,clip]{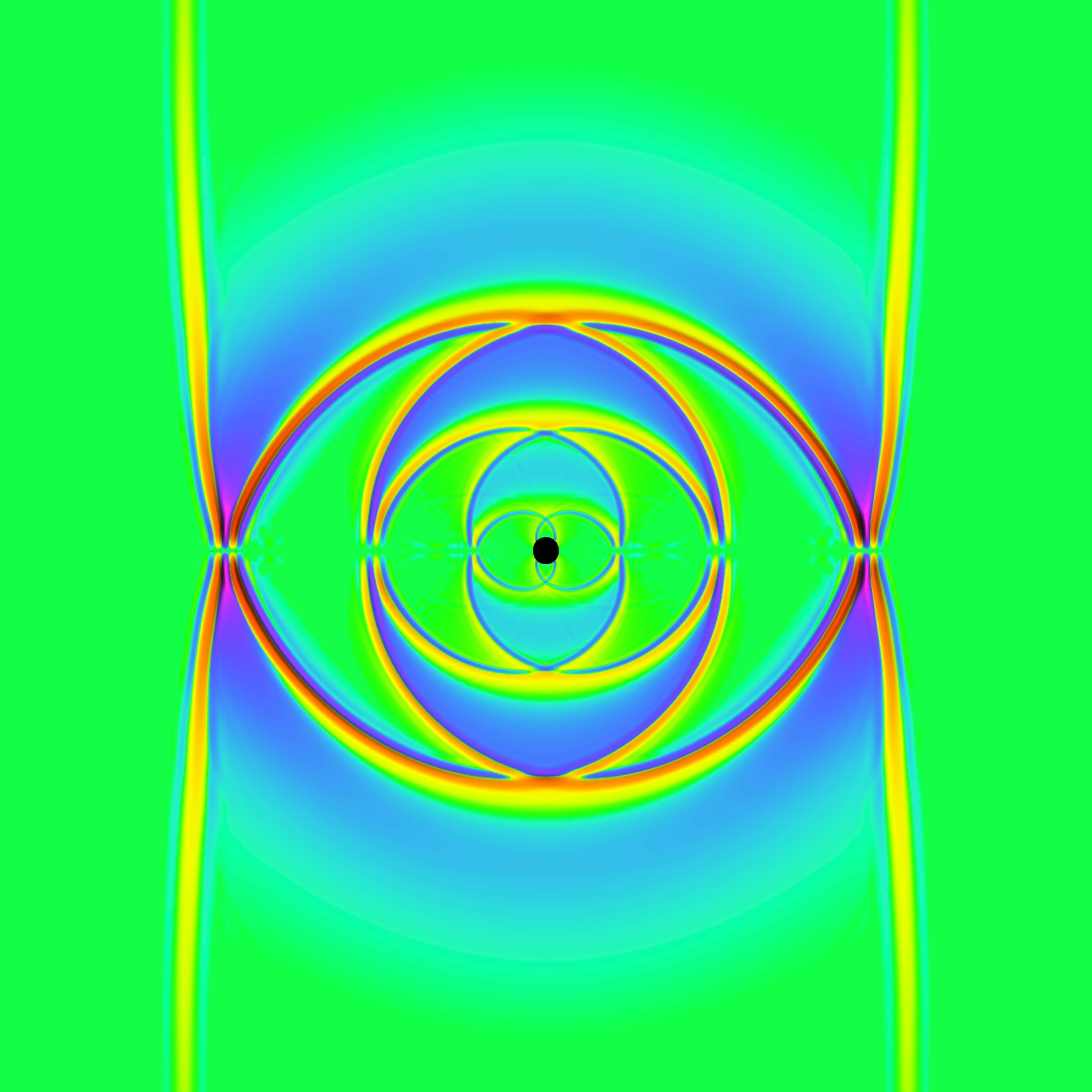} 
\end{center}
\caption{
Four snapshots of $^{(r)}\Psi_4 \times r M$ from the collision
described in the text. Times from top left to bottom right
are $t=0M, 5M, 20M$, and $50M$. The symmetry ($x$) axis is horizontal and passes through
the center of each panel; we only simulate the top ($y\ge 0$) half of the plane,
but for visualization purposes also show $y\le 0$. The size of each panel
is $150M\times150M$. 
The central black dots in the panels after $t=0$ indicate the black hole
that formed, and are specifically the excised regions, set to $60\%$ the size of 
the apparent horizon at each time. 
By $t=5M$, $>99.9\%$ of the null particles sourcing the initial wave fronts 
have fallen into the black hole, and to excellent approximation then and in subsequent
panels the spacetime is vacuum exterior to the excised regions.
\\
(Note that the implied amplitude of the initial wave fronts
in the top left panel is misleading. At $t=0$, the two pulses are moving in the $x$ direction,
but are offset from $y=0$, so that $^{(r)}\Psi_4$ measures
some radiation is simply because the $r$ tetrad direction has some overlap 
with the propagation direction. However, that the {\em magnitude} appears not 
to decay with distance from origin at $t=0$ is entirely due to resolution effects
( $^{(x)}\Psi_4$ given by (\ref{xpsi4}) drops off like $1/\rho^2$): the mesh-refinement
algorithm successively lowers the resolution moving outward, and because
of how thin in $x$ the initial waves are, far from the origin these
features are significantly under-resolved. This is exacerbated in the calculation
of $^{(r)}\Psi_4$, which requires second gradients of the metric. As evolution
proceeds, the Kreiss-Oliger numerical dissipation we use smooths out these
under-resolved features.)
}
\label{fig_psi4}
\end{figure*}

See Fig.\ref{fig_psi4} for a plot illustrating the collision in terms
of the gravitational radiation produced,
measured with $^{(r)}\Psi_4$, the Newman-Penrose scalar 
with tetrad adapted to measure outgoing radiation propagating in the radial coordinate
direction $r=\sqrt{x^2+y^2+z^2}$. 
Evident in the figure is that the radiation, roughly speaking, appears to be composed
of two components. One is the longer wavelength feature expected from
the dominant quadrupolar quasi-normal mode (QNM) oscillation of the black hole
created during the collision. The other 
can be associated with the initial wave front, the inner part of which is trapped
by the black hole, while the remainder propagates about
the black hole {\em ad infinitum}, forming the concentric, circularlike features
of characteristic wavelength close to $2\Delta u$. In the AS limit where $\Delta u\rightarrow 0$,
the leading edge of this feature presumably remains shocklike.

The tetrad used to compute $^{(r)}\Psi_4$ is completed with vectors
tangent to the sphere $r={\rm constant}$; we define the corresponding angles
on the sphere so that $\theta$ measures the angle from the $z$ axis (so the simulation
plane $z=0,y\ge0$ is $\theta=\pi/2$), $\phi=0$ ($\phi=\pi$) coincides with the plane
$x>0,y=0$ ($x<0,y=0$), and $\phi=\pi/2$ ($\phi=3\pi/2$) coincides with the plane
$x=0,y>0$ ($x=0,y<0$). Thus, on the positive $x$ axis, this tetrad will
be exactly that used to define $^{(x)}\Psi_4$ (\ref{xpsi4}), and similarly
on the negative $x$ axis but flipped to measuring radiation propagating outward along
the $-x$ direction. 
This then gives us one way to easily understand the
feature in Fig.~\ref{fig_psi4} that $^{(r)}\Psi_4$ vanishes on the collision $(x)$ axis.
Even though the presence of the matter forces 
$^{(x)}\Psi_4$ to vanish on the axis in the initial data [see the discussion
after equation (\ref{xpsi4})], most of the matter region immediately
falls into the black hole, and essentially all the radiation that reaches the
axis can be traced back to vacuum regions along the initial wave fronts. Here, the
polarization relative to an $x$ direction propagation
vector [indicated by the $\cos(2\theta)$ and $\sin(2\theta)$ terms
in (\ref{xpsi4})] required for the wave to be axisymmetric about the collision
axis is such that when a ring of waves focuses to the axis, their sum must be zero.

\begin{figure}
\vspace{0.1in}
\includegraphics[width=3.00in,draft=false,clip]{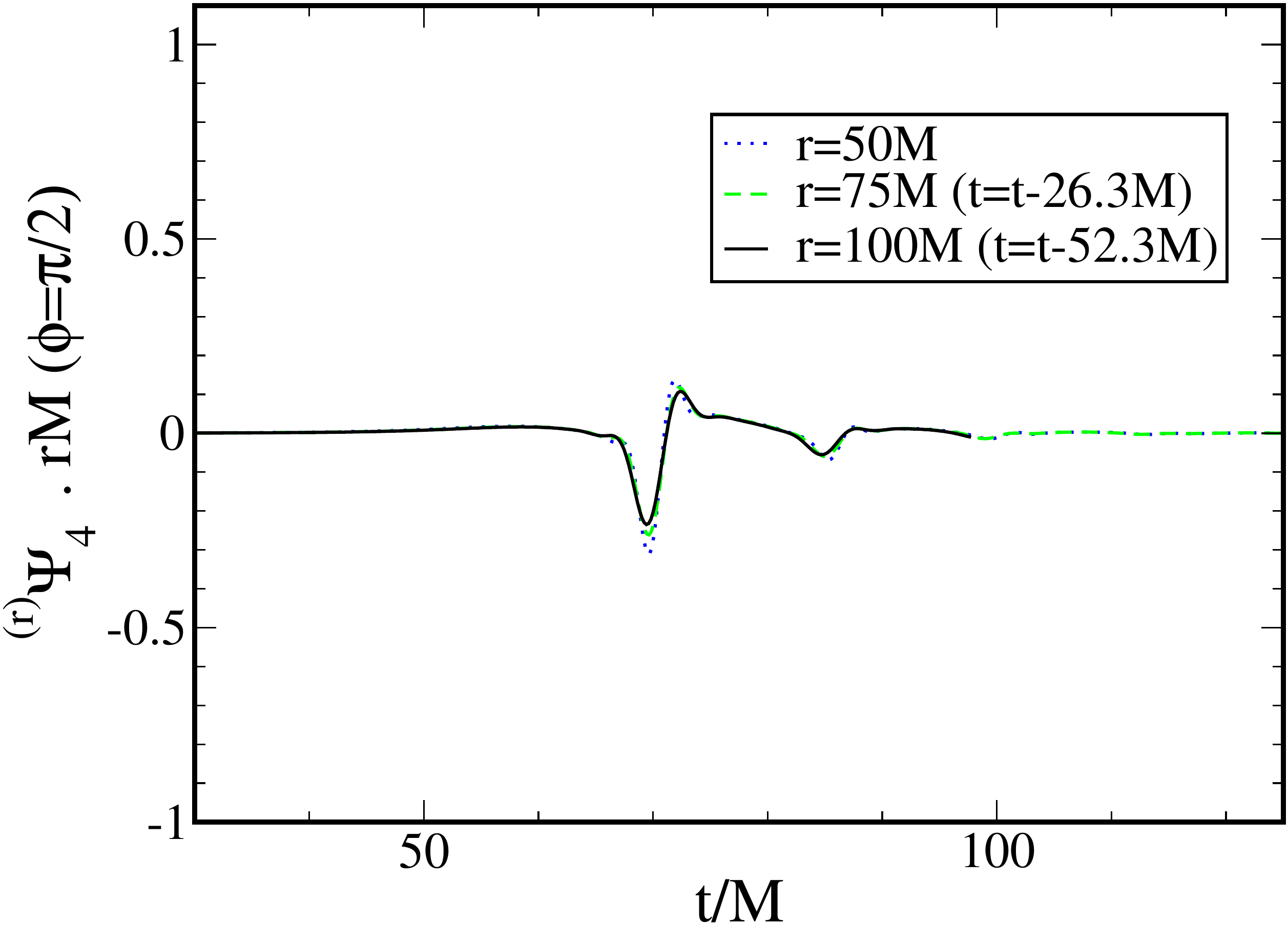}
\includegraphics[width=3.00in,draft=false,clip]{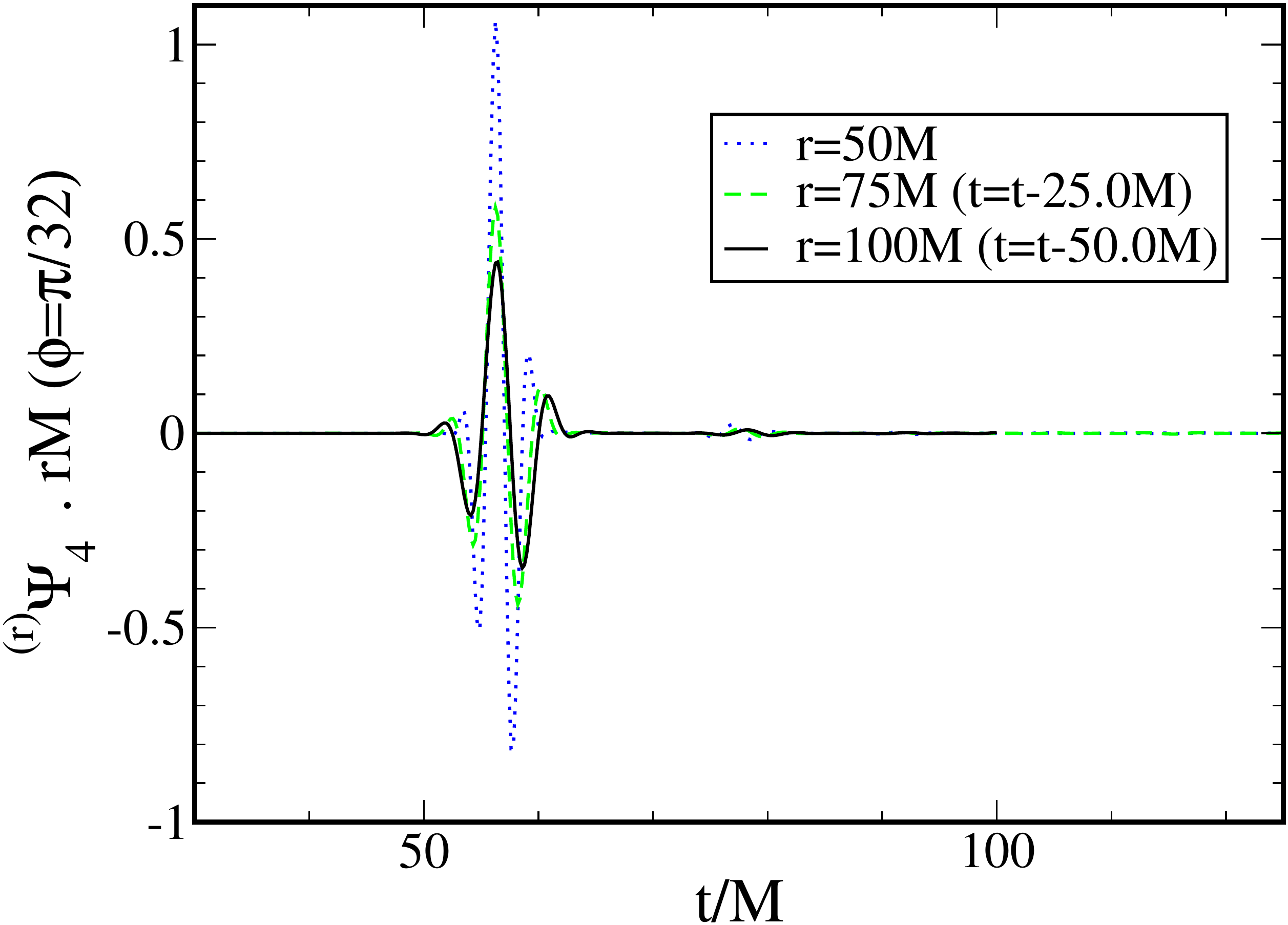}
\caption
{
$^{(r)}\Psi_4(t) \times rM$ measured at three radii and two angles 
relative to the origin (the top panel is at a right angle
relative to the collision axis, the bottom panel is at
$\phi=\pi/32$ from the collision axis); see Fig.~\ref{fig_psi4_cuts_ln_abs}
for the same data on a logarithmic vertical scale.
The $r=75M$ and $r=100M$ date were shifted in time by the amounts shown
in the legend to account for the different propagation times (and note
that these shifts are different for the two panels). 
Though the simulation was run to $t=175M$, the data for the $r=100M$
point was truncated at $t=150M$ to avoid possible artifacts
coming from the computational boundaries at $|x|=y=250M$, assuming
a unit coordinate light speed. Note that, particularly near the
axis, we are not yet sufficiently far from the collision that
we see the expected $1/r$ decay of the wave. (In the wave zone, given the scaling
    of $\Psi_4$ by $rM$, the shifted waves should have the same amplitudes.)
}
\label{fig_psi4_cuts}
\end{figure}

\begin{figure}
\vspace{0.1in}
\includegraphics[width=3.00in,draft=false,clip]{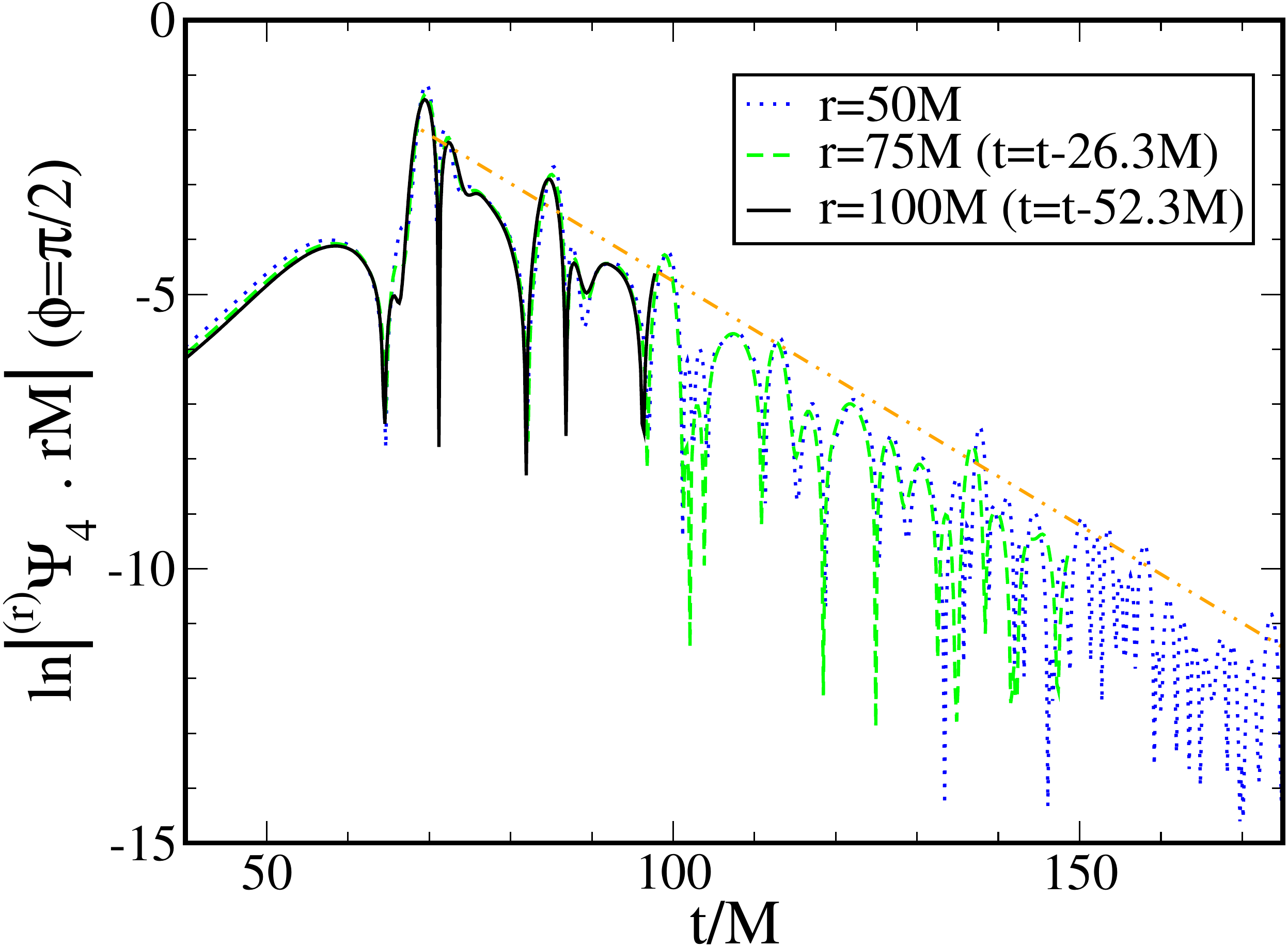}
\includegraphics[width=3.00in,draft=false,clip]{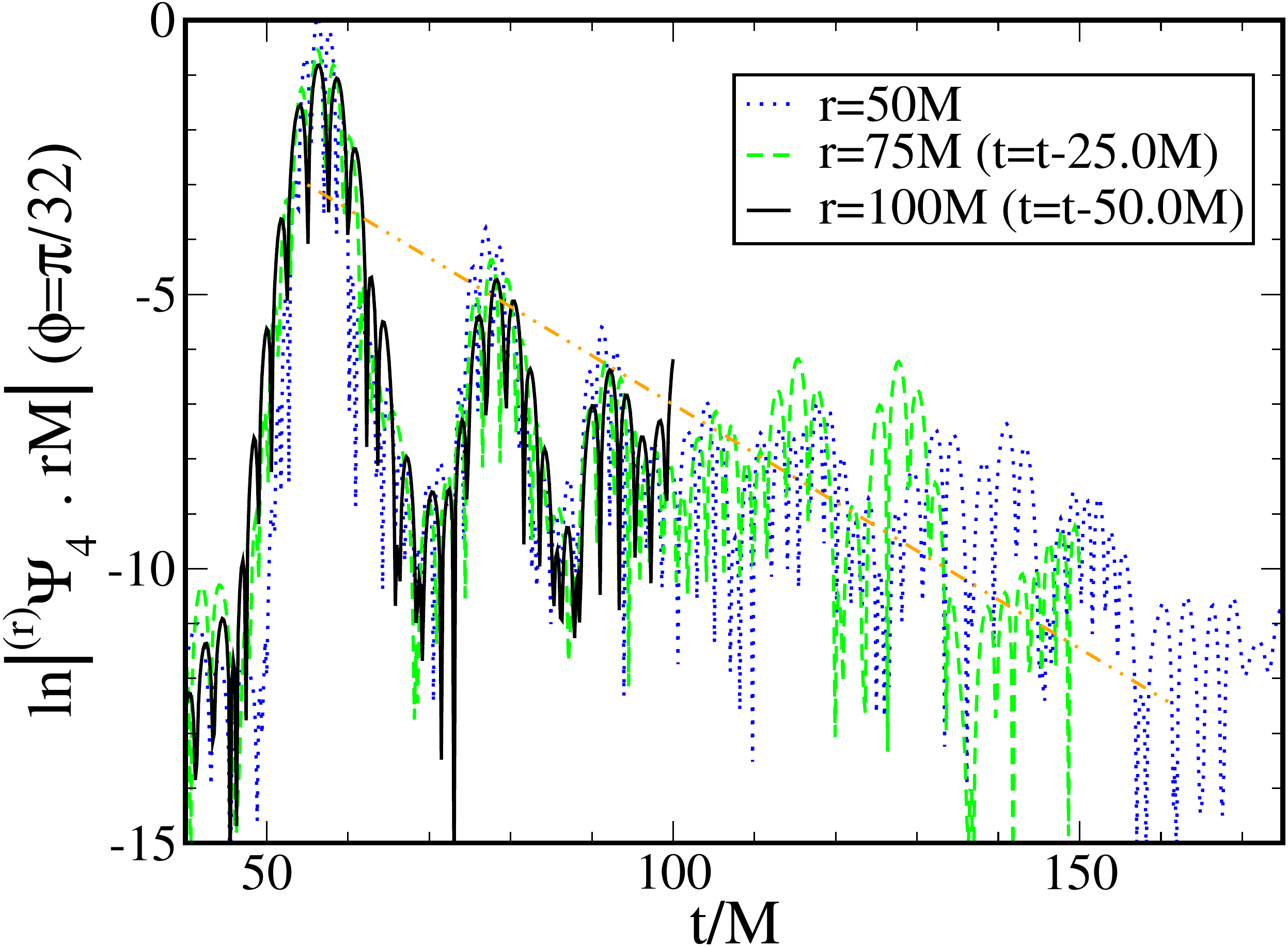}
\caption
{The natural logarithm of the magnitude of the same data shown in Fig.~\ref{fig_psi4_cuts}.
For reference, overlayed (orange double-dot dash) is a straight line segment
with slope $-0.089$, which is the expected QNM decay rate of the
least damped $\ell=2$ quadrupolar mode (see e.g.~\cite{Berti:2005ys});
note that this line is {\em not} a fit to the data.
}
\label{fig_psi4_cuts_ln_abs}
\end{figure}

Despite the fact that symmetry forces $^{(r)}\Psi_4$ to be exactly zero on the axis,
an interesting property of the radiation is how strongly beamed
it is about the axis (this is consistent with perturbative calculations of
the scattering problem, and also noted in high speed collisions of black
holes~\cite{Sperhake:2008ga,Healy:2015mla}). This is further illustrated
in Fig.~\ref{fig_psi4_cuts} showing $^{(r)}\Psi_4 (t)$ at three radii
($r=50M$, $75M$, and $100M$), and two angles relative to the collision axis 
(at a right angle $\phi=\pi/2$, and close to it with $\phi=\pi/32$). Figure~\ref{fig_psi4_cuts_ln_abs}
is the same data plotted on a logarithmic scale, to highlight the
decay of the waves. Interestingly, the latter plot shows that, transverse
to the collision axis, the decay rate is broadly consistent with the
decay of the least damped QNMs, but less so near the axis.
Also, the time shifts required to align the waves on the
plots indicate different effective propagation
speeds along the axis versus transverse to it; how much of this
is simply due to gauge (time slicing, how the coordinate $r$ relates to
some geometric radius, etc.) as opposed to slightly different geometries
the different parts of the wave are propagating through is unclear.

The above observations suggest at least part of the outgoing radiation might be not be
captured as a sum of Schwarzschild QNMs (which do not form a complete basis),
as in a sense the initial wave-fronts are part of the initial data, and not
``produced'' by a perturbed black hole. However, since the decay of such a wave-front
propagating about the black hole will be controlled by the unstable photon orbit,
similar to QNMs, at least at late times there may be no practical distinction between 
what is a QNM versus what is a remnant of waves from the initial data.

\begin{figure}
\vspace{0.1in}
\includegraphics[width=3.00in,draft=false,clip]{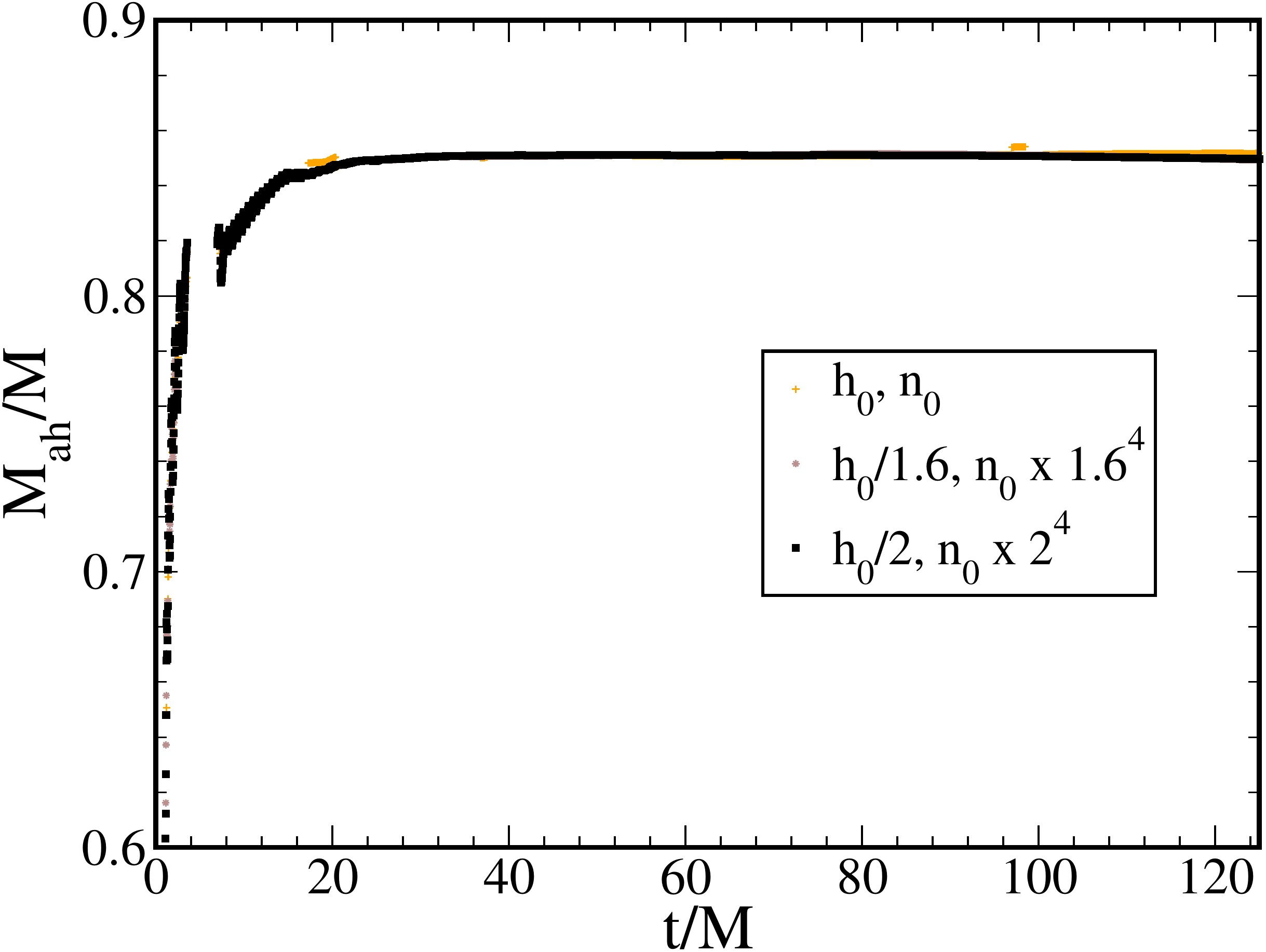}
\caption
{The mass of the black hole estimated from the area of the apparent
horizon, for three different resolutions. The apparent horizon
is first detected around $t=1.1M$ in all cases, though between
$t\sim4-7M$ the apparent horizon finder fails to find it to within a reasonable
tolerance (we use a flow method, assuming a ``star shaped'' surface, and though
the horizon is quite deformed here, it is not close to violating this 
assumption). We are not sure why this happens, however, the temporary
loss of this surface that guides excision (the excision surface is set to
the same shape as the apparent horizon, but $60\%$ its  size, and is frozen in 
shape when the finder fails) does not affect stability
at the excision surface, implying characteristics remain ingoing there.
}
\label{fig_ah_m}
\end{figure}

\begin{figure}
\vspace{0.1in}
\includegraphics[width=3.00in,draft=false,clip]{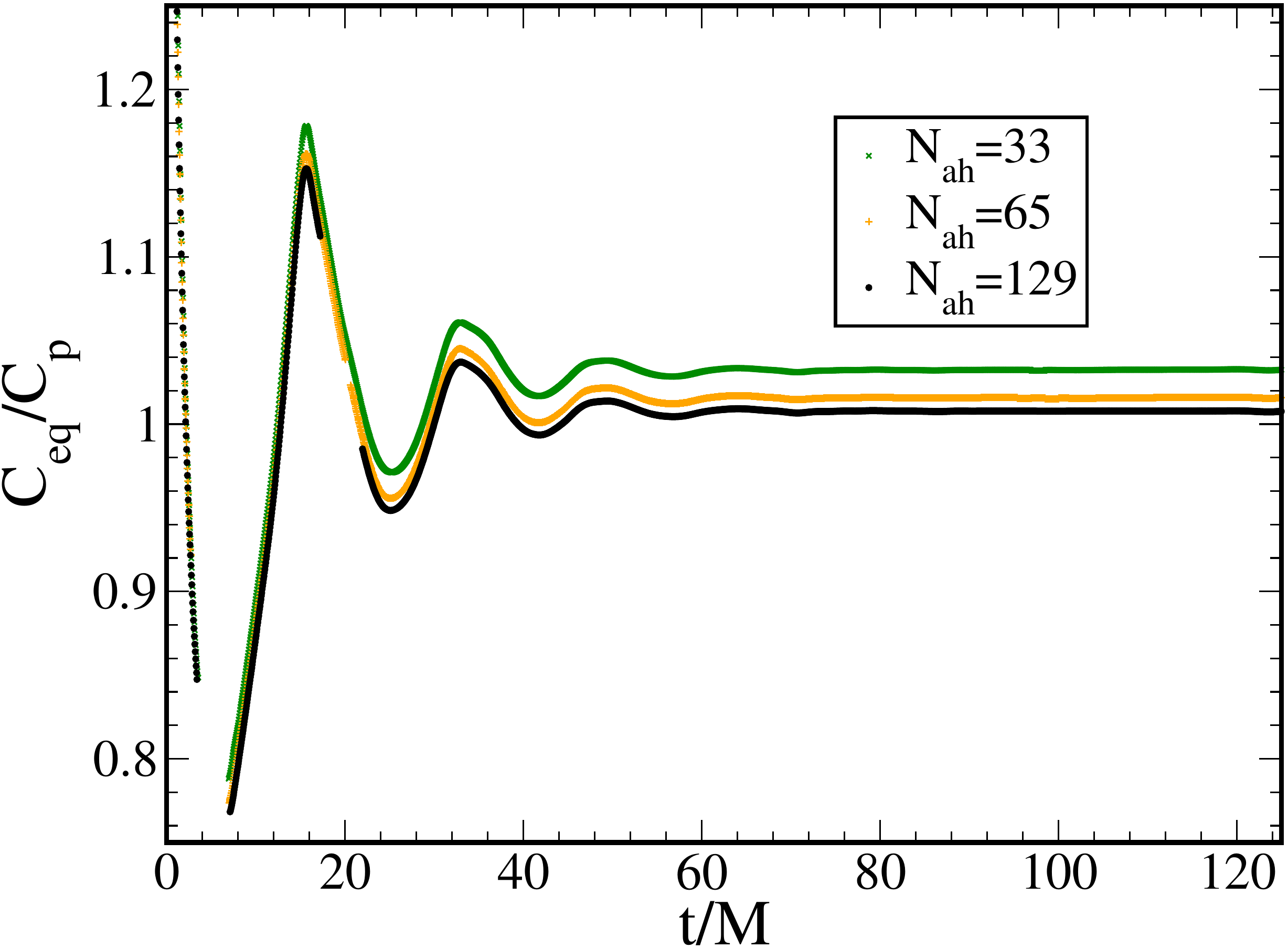}
\caption
{
The ratio of proper equatorial to proper polar circumference $C_{eq}/C_p$ of the 
apparent horizon from the $h_0$ resolution case, to illustrate the initial dynamics of the horizon
and its ringdown to a Schwarzschild black hole. For the default
resolution we use to resolve the horizon, namely uniformly
discretized in $\phi$ with $N_{ah}=33$ points, $C_{eq}/C_p$ is not
calculated with a similar accuracy as the area; see Fig.~\ref{fig_ah_m}
(for which the estimated fractional error in $M_{ah}$ is $0.8\%$,
compared to $3.3\%$ for $C_{eq}/C_p$). However, this is not a reflection
of the underlying accuracy of the spacetime solution, as demonstrated
by the higher resolution apparent horizon finder curves, with
the {\em same} $h_0$ spacetime resolution. $C_{eq}/C_p$ is converging
to the expected value of one at $1^{st}$ order 
[dominated by integration truncation error on the axis ($\phi=0,\pi$)]. 
We do not typically run with higher apparent horizon resolutions,
as the flow method's computation time scales poorly with $N_{ah}$,
and is also more challenging to robustly find a very dynamical horizon
to within low tolerance, as indicated by additional times
between $t\sim18-22M$ when the higher resolution cases fail to find
the horizon.
}
\label{fig_ah_cp}
\end{figure}

In Fig.~\ref{fig_ah_m} we show an estimate of the black hole mass
that forms as a function of time, calculated by computing
the area of the apparent horizon. In Fig.~\ref{fig_ah_cp} we show
the ratio $C_{eq}/C_p$ of proper equatorial to polar circumference
of the horizon, illustrating the early time dynamics and subsequent
decay to a Schwarzschild black hole. From the area, we estimate
the mass of the remnant black hole to be $M_{ah}=0.851\pm0.007M$. Given that 
the particles that escaped being trapped by the black hole
collectively contain less than $0.001M$ of energy, we can infer that
the energy emitted in gravitational waves is $E_{\rm GW}=14.9\pm 0.8\%$ 
of the initial spacetime mass $M$. 

Integration of $^{(r)}\Psi_4$ measured on coordinate
spheres gives order of magnitude consistent
values, specifically $20.3\%$, $18.0\%$, and $16.3\%$ integrated
on spheres of radius $r=50M$, $75M$, and $100M$,
respectively, from the $h_0/2$ run. However, this does not 
seem to be sufficiently far into the wave zone, in that we
do not yet see the expected $1/r$ decay in the waveform, which 
would allow extrapolation to
$r=\infty$ for an accurate estimate. This mostly
seems to be due to near-axis beaming of the radiation, illustrated 
in Fig.~\ref{fig_psi4_cuts}. In addition, a couple of other
effects, one gauge, the other numerical, hinder 
trying to fit the energy to a more complicated $1/r$ series
expansion, and so for now we will take the apparent horizon
based estimate as the more accurate measure of total energy radiated.
The numerical
issue is related to the short wavelength component of the radiation,
proportional to the width $2 \Delta u$ of the initial data,
and is responsible for the concentric rings evident in Fig.~\ref{fig_psi4}.
This is not well resolved by the grid in the wave extraction zone,
and numerical dissipation attenuates this feature more
rapidly than the longer wavelength components of the wave.
The gauge issue is related to the significant gauge dynamics
we have as we transition from the initial data coordinates to the damped harmonic
gauge post collision. This is more pronounced closer to the
origin, and is exacerbated by our naive procedure of measuring
radiation on $r=\rm{constant}$ coordinate spheres, and simply
using $4\pi r^2$ as their geometric area (which is only correct
asymptotically). Certainly a combination of improved resolution,
farther extraction, and a more geometrically sound construction
of extraction spheres could alleviate these issues, though that will
take considerable effort and computational resources, and we leave it
to future work.

\section{Conclusions}\label{sec_conc}

We have described a formalism for studying the ultrarelativistic
scattering problem using plane-fronted distributions of null
particles as a matter source. We have developed a numerical code 
based on this, and as a first application presented a study of black hole
formation in head-on axisymmetric collisions, with parameters
of the particle sources chosen to give a postcollision spacetime
close to that expected to be produced by the collision
of two Aichelburg-Sexl shock waves. We find results broadly
consistent with prior studies of this limit, whether perturbatively,
or via full numerical solution but using finite boost compact 
object models of particles. Specifically, based on the area of the
resultant black hole that forms, we infer $14.9\pm0.8\%$ of the initial
mass of the spacetime is radiated as gravitational waves during the
collision. We leave it to future work to explore in detail where
this particular case fits into a limiting sequence
reaching the AS limit, though we estimate in terms of net gravitational
wave emission this is slightly below the AS limit by $\sim0.3\%$ 
(hence the quoted value of $15\pm 1\%$ in the abstract as the implied
value for the AS limit).

In terms of other future directions, there are numerous avenues that
can be pursued, many of them outlined in the introduction,
so we will not repeat that discussion here. Rather, we will briefly
mention a few outstanding issues in the code that would
need to be addressed before the full breadth of applications
could be tackled. First is simply the computational expense
of the simulations, in part due to the scaling 
of particle number $N$ required to achieve a desired
level of accuracy (which will be more severe for 3D applications
than the axisymmetric 2D case presented here), and in part that the collisionless
particle 
model generically allows focusing to caustic regions. In principle,
the latter problem could be alleviated by including some form
of self-interaction between the particles that produces
an effective pressure; however, it is unclear to what extent that kind
of matter could be used as a consistent source for plane-fronted
gravitational wave spacetimes.
A second issue is that earlier studies (in particular~\cite{East:2012mb}),
and preliminary investigations with this code, show the gauge conditions
(i.e. source function evolution equations) that work well for prompt black hole
formation are not adequate for slightly weaker interactions where the spacetime
is highly dynamical but does not immediately (or will not ever) form horizons. What typically
happens with the current gauges is a coordinate singularity forms. 
This is also a problem that has plagued attempts to study vacuum critical
collapse (see e.g.~\cite{Hilditch:2017dnw}), and if the critical solution
is universal it would not be surprising
if a single class of novel gauge condition could solve the coordinate issues
for both these applications.

\acknowledgments %
We thank Herbert Balasin and Peter Aichelburg for useful comments
regarding an earlier version of this manuscript.
F.P.  acknowledges support from NSF grant PHY-1607449, the Simons
Foundation, and the Canadian Institute For Advanced Research (CIFAR).
This research was further supported in part by the Perimeter
Institute for Theoretical Physics. Research at the Perimeter Institute is
supported by the Government of Canada through the Department of
Innovation, Science and Economic Development Canada and by the
Province of Ontario through the Ministry of Research, Innovation and
Science.  Computational resources were provided by XSEDE under grant
TG-PHY100053 and the Perseus cluster at Princeton University.

\appendix
\section*{Appendix: Application to Inhomogeneous Cold Matter Cosmology}
In this appendix, we briefly present some results of applying the methods
discussed in the main text for evolving the Einstein equations, coupled to
particle distributions, to the case of inhomogeneous cold matter in an expanding
universe.  We do this both to demonstrate that these same methods can be
applied to massive (timelike) particles, and in three spatial dimensions, as well
as to further validate the code by comparing it to known results. In
particular, we repeat a calculation from~\cite{East:2017qmk} where we start with
a (timelike) dust-filled, expanding universe (i.e. the Einstein-de Sitter model) and include
some initially small inhomogeneities that are all at a wavelength that is four times the
initial Hubble radius, and with velocity given by the Zel'dovich approximation. 
(In particular, this is the case labeled
$\bar{\delta}=10^{-3}$ in~\cite{East:2017qmk}; see that reference for details.)  

To construct initial data, we begin with a solution to the constraint equations
obtained using the code of~\cite{idsolve_paper}, as described in~\cite{East:2017qmk},
which uses a fluid description of the matter. We then create a particle
distribution that is consistent with the density field in the following manner.
We begin with a uniformly spaced lattice of particles that we then perturb from their
positions according to the Zel'dovich approximation, as is typically
done in N-body calculations. In particular, we apply the shift in position given
by Eq.~(31) in~\cite{Chisari:2011iq}.
We assign the velocity of the particle by
interpolating the fluid velocity field to the particle position. 

By default the particles would have uniform masses given by $m_i=\rho_0L^3/N$,
where $\rho_0$ is the initial homogeneous density, $L^3$ is the volume of the domain,
and $N$ is the number of particles.
However, we then
calculate a small, nonlinear correction to the mass of each particle by
first finding the density given by $\bar{\rho}_e=-T_a^a$, with $T^{ab}$
calculated from (\ref{set_average}) using uniform particle masses, and 
taking the ratio with the
desired density field from the fluid representation at each particle position $\rho_e/\bar{\rho}$.
We then rescale the mass of each particle by this factor 
$m_i\to m_i\times(\rho_e/\bar{\rho})$.

\begin{figure}
\vspace{0.1in}
    \includegraphics[width=\columnwidth,draft=false,clip]{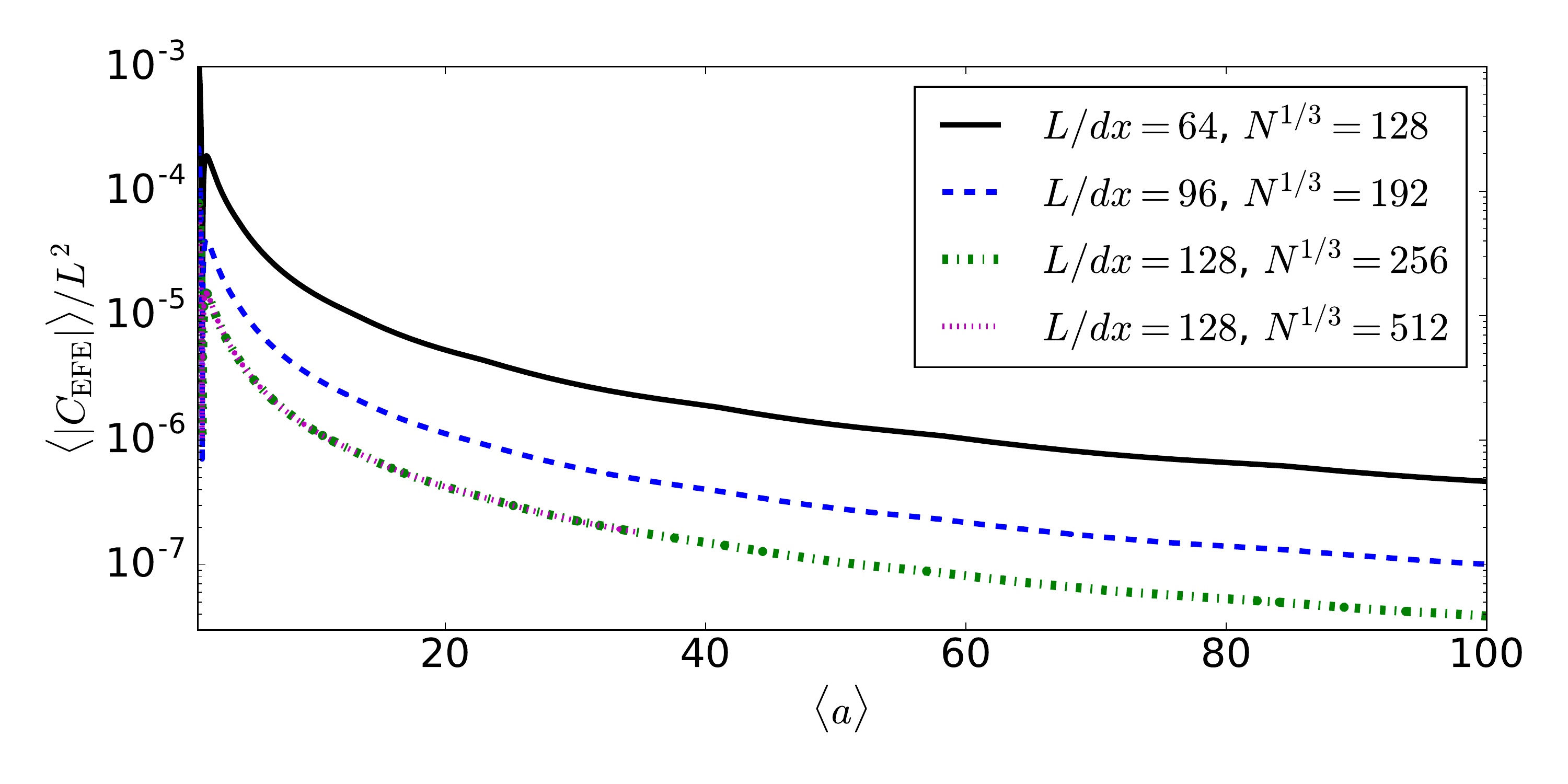}
\caption
{
The volume average of the L2-norm of the constraints (\ref{harm_const}) over
    the domain at several different spatial resolutions and particle numbers.
    At these resolutions, the constraint violation is dominated by the spatial
    discretization truncation error, as opposed to that coming from the finite
    number of particles (the $N^{1/3}=512$ case is not continued
    as long as the others for this reason). The decrease in the constraint violation with
    increasing spatial resolution is faster than $2^{nd}$ order convergence,
    suggesting that the error from reconstructing the stress-energy tensor is subdominant
    to that coming from evolving the metric and particles.
}
\label{fig_cosmo_cnst}
\end{figure}

The evolution is performed as described in the main text, except that our
domain is three-dimensional and periodic (particles that exit the domain at one 
boundary are wrapped around to the opposite boundary), and the particles have nonzero rest
mass.  In Fig.~\ref{fig_cosmo_cnst}, we demonstrate the convergence of the
constraints (\ref{harm_const}) by performing the calculation at several resolutions
ranging from $64^3$ to $128^3$ grid points covering the domain.
As in the main text, the error from finite particle number is subdominant to the grid
discretization error for the parameters considered here.

\begin{figure}
\vspace{0.1in}
    \includegraphics[width=\columnwidth,draft=false,clip]{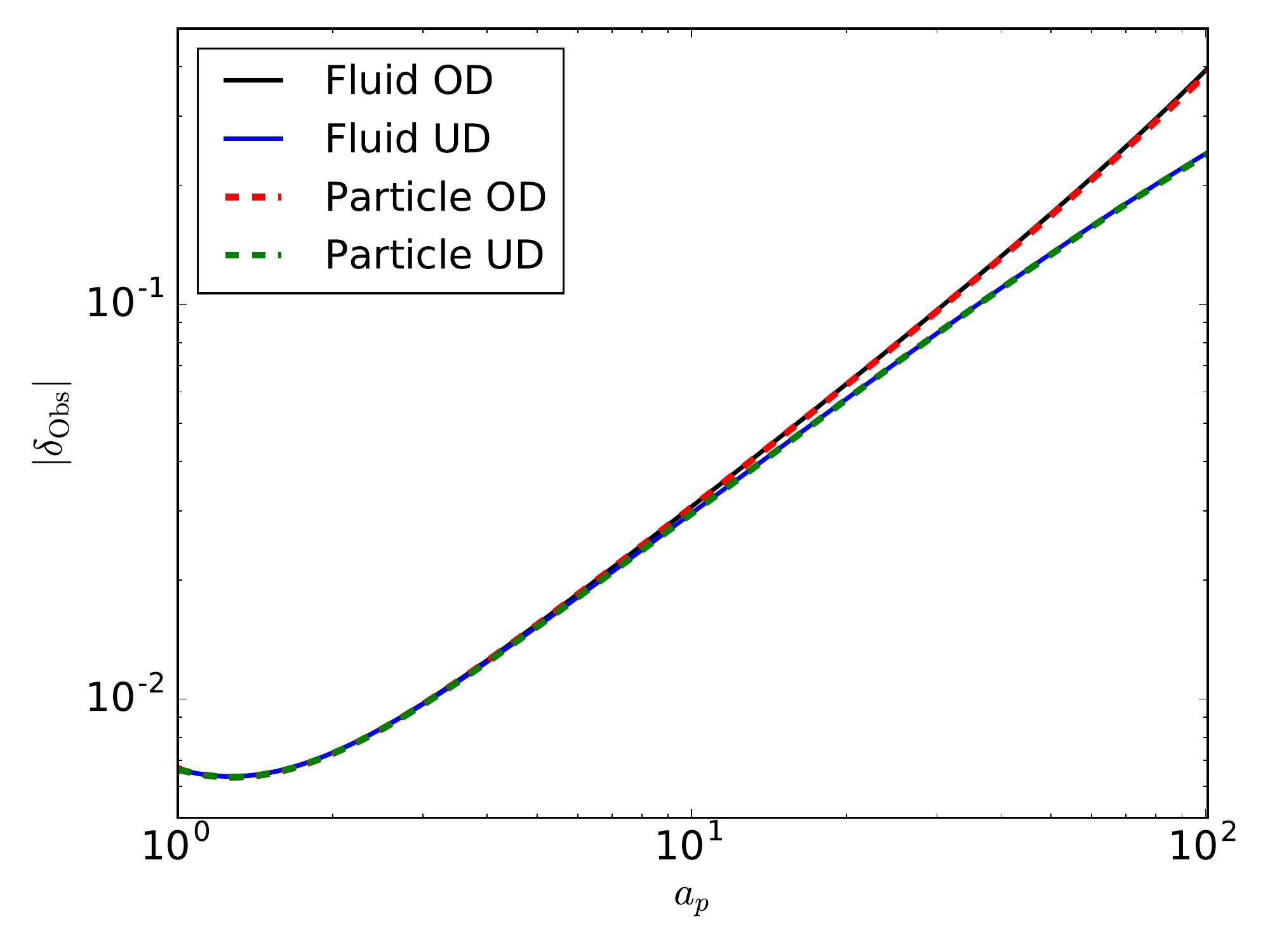}
\caption
{
Comparison of the density contrast measured by an observer comoving with the
matter at the points of maximum (labeled OD) and minimum (labeled UD)
density versus a local measure of the scale factor (see~\cite{East:2017qmk}
for exact definition).
The results obtained with particle evolution closely track those obtained in~\cite{East:2017qmk}
by evolving the fluid equations.
}
\label{fig_cosmo_density}
\end{figure}

We can also compare the evolution of the density contrast, which increases and becomes
nonlinear (as signaled by
the divergence of the under and overdensities) 
as the inhomogeneities enter the horizon, to the results from~\cite{East:2017qmk},
which were obtained by evolving a pressureless fluid. As long as we restrict to times
before multistream regions form, the two treatments should give the same answer.
As shown in Fig.~\ref{fig_cosmo_density}, the density contrast from the two cases
is indeed a good match.

The advantage of the particle treatment is that it allows one to evolve through the formation
of multistream regions that will arise during structure formation, and thus be more directly
comparable to Newtonian N-body simulations. However, we leave a study of this to future work.

\bibliography{paper1}\label{refs}

\end{document}